\documentclass[12pt,notitlepage,nofootinbib]{book}

\usepackage{graphicx}
\usepackage{amsmath}
\usepackage{amssymb}
\usepackage{dcolumn}
\usepackage{bm}
\usepackage{color}
\usepackage{dsfont}

\newcommand \beq{\begin{eqnarray}}
\newcommand \eeq{\end{eqnarray}}

\newcommand \be{\begin{equation}}
\newcommand \ee{\end{equation}}
\newcommand{\nn}{\nonumber \\}

\newcommand{\eq}[1]{\begin{align} #1 \end{align}}

\title{\Large \bf THREE LECTURES ON THE CURCI-FERRARI MODEL}

\author{Urko Reinosa}

\date{\it Centre de Physique Th\'eorique, CNRS, Ecole polytechnique,\\ IP Paris, F-91128 Palaiseau, France.}

\begin{document}

\maketitle

\tableofcontents

\chapter*{Acknowledgements}
I warmly thank the IFT and the ICTP/SAIFR at S\~ao Paulo for their hospitality during my two week stay in Brazil where these lectures were given. Special thanks to Gast\~ao Krein and Duifje van Egmond for making sure that everything went smoothly and also to the students, postdocs and some seniors for the many interactions.

\chapter*{Introduction}

The Curci-Ferrari (CF) model is an approach to Landau gauge-fixed Yang-Mills (YM) theory and/or Quantum Chromodynamics (QCD) that aims at capturing phenomenologically the effect of the Gribov copies. The latter usually spoil the validity of the Faddeev-Popov (FP) gauge fixing procedure beyond the ultraviolet (UV) range and call for the construction of new gauge-fixed actions that could help investigating the infrared (IR) regime of these theories in the continuum.\footnote{Lattice simulations of these theories can be done without fixing the gauge. They can also be done within a gauge-fixed setting, for certain gauges.} To date, unfortunately, a fully consistent gauge-fixed action in the IR range is not known. One can try to model this ignorance by adding new operators to the FP action and try to constrain their couplings (or in some cases discard them) by using experimental data or the results of lattice simulations.

\begin{figure}[t]
\begin{center}
\includegraphics[height=0.25\textheight]{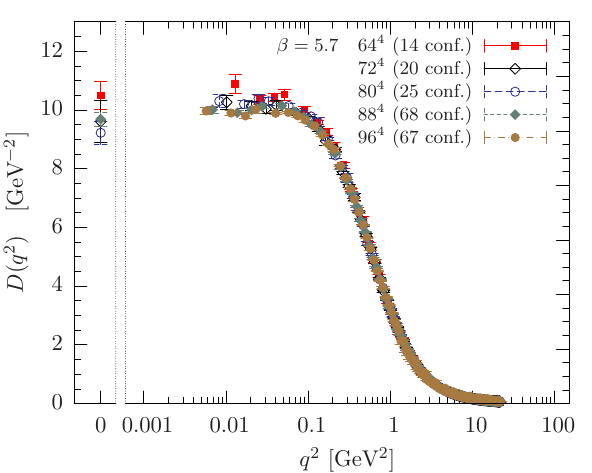}\includegraphics[height=0.25\textheight]{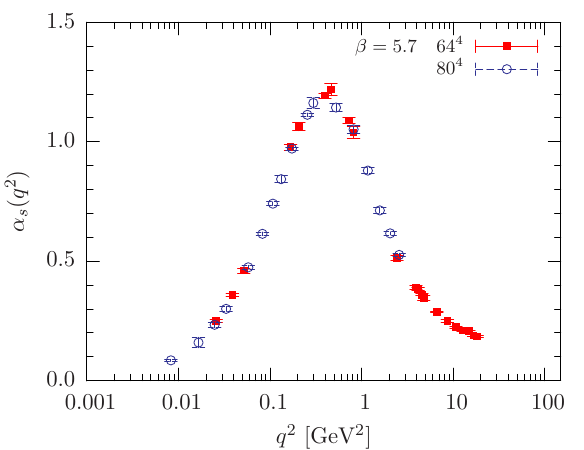}\\
\caption{Left: the Landau gauge Euclidean gluon propagator as computed on the lattice \cite{Bogolubsky:2009dc}. Right: the Landau gauge running coupling in the Taylor scheme as compute on the lattice, in the Euclidean domain \cite{Bogolubsky:2009dc}. Note that the perturbative expansion parameter is $g^2N_c/16\pi^2=\alpha_SN_c/4\pi$ which does not exceed $1$ in the present plot.}\label{fig:lattice}
\end{center}
\end{figure}

In particular, over the past decades, lattice simulations have revealed surprising results on the way the degrees of freedom in the Landau gauge behave/are coupled to each other in the Euclidean IR domain. Notably, the gluon two-point function is seen to saturate to a finite non-zero value at small momenta akin to a screening mass, see Fig.~\ref{fig:lattice}. Moreover, in the case of pure Yang-Mills theories, the coupling constant does not display the usual Landau pole behavior found in the FP approach as the running scale is decreased, but, rather, saturates at certain scale and even decreases again as the running scale is decreased further. A the same time, the maximal value reached by this coupling is not large (in the YM case), see Fig.~\ref{fig:lattice}.

This poses the question of whether a proper account of the Gribov ambiguity could lead to a gauge-fixed theory which could admit a well defined perturbation theory to all scales and which would then allow one to capture some of the aspects of Euclidean YM theory in the IR. Another question is whether an effective way to capture the effect the Gribov copies could be to incorporate a gluon mass term to the incomplete Faddeev-Popov action, as suggested by the lattice simulations. These are some the questions that are investigated within the CF set-up.

In these lectures, we review to which extent the CF model allows one to capture perturbatively some of the properties of Landau-gauge YM theories in the Euclidean domain. We limit ourselves to the two-point correlation functions, as originally discussed in Refs.~\cite{Tissier:2010ts} and \cite{Tissier:2011ey}. These objects are of course not observable but comparing them to the correlators evaluated on the lattice provides a good test of the CF model as an effective description of gauge fixing in the IR. Applications to the evaluation of certain observables that can be defined in the Euclidean domain will be discussed in the conclusions section. Extensions to QCD and/or beyond the Euclidean range will also be briefly mentioned there.

In the first lecture, we shall introduce the model and derive some of its properties, in particular two useful non-renormalization theorems. We provide a new and simpler derivation of the non-renormalization theorem associated the CF mass. The second lecture will address the evaluation of the two-point correlators at one-loop order and the comparison to the lattice data. Finally, in the third lecture, we shall perform a renormalization group analysis which tells in particular what is the fate of the model at large and low momentum scales. For reviews on recent applications of the CF model, see Refs.~\cite{Pelaez:2021tpq} and \cite{Reinosa:2019xqq}.

\chapter{General Properties}
The Curci-Ferrari (CF) model is defined by the Euclidean action
\beq
S=\int_x \left\{\frac{1}{4}F_{\mu\nu}^aF_{\mu\nu}^a+\partial_\mu\bar c^a D_\mu c^a+ih^a\partial_\mu A_\mu^a+\frac{1}{2}m^2A_\mu^a A_\mu^a\right\},\label{eq:CF}
\eeq
with $\smash{\int_x\equiv \int d^dx}$. The non-integer dimension $\smash{d=4-2\epsilon}$ is used to regulate the UV divergences that appear upon evaluating quantum fluctuations.\footnote{Example of those will be discussed in the next lecture.}

The first term of Eq.~(\ref{eq:CF}) is the (Euclidean) Yang-Mills action. The next two terms correspond to the gauge fixing in the Landau gauge $\smash{\partial_\mu A_\mu^a=0}$ within the Faddeev-Popov framework, which is valid only at large momenta in principle, as we have recalled in the Introduction. The last term is a (lattice motivated) phenomenological input meant to account, in part, for the deficiencies of the Faddeev-Popov paradigm at low momenta. We have used the widespread notations
\beq
F_{\mu\nu}^a\equiv\partial_\mu A_\nu^a-\partial_\nu A_\mu^a+gf^{abc}A_\mu^b A_\nu^c\,,
\eeq
for the non-Abelian field-strength tensor, and
\beq
D_\mu\varphi^a\equiv\partial_\mu\varphi^a+gf^{abc}A_\mu^b\varphi^c\,,
\eeq
for the adjoint covariant derivative.

In this lecture, we discuss some of the general properties of the Curci-Ferrari model encoded in the corresponding correlation/vertex functions.

\pagebreak

\section{A First Look at the Feynman Rules}
The Feynman rules are the building blocks of Feynman diagrams that allow one to organize perturbative calculations. They can also be used to infer some general properties obeyed by the correlation/vertex functions, valid to all orders.

\subsection{Free Gluon Propagator}
The free gluon propagator is defined as
\beq
G_{\mu\nu}^{ab}(x,y)\equiv\frac{\int {\cal D}[Ac\bar ch]\,A^a_\mu(x)A^b_\nu(y)\,e^{-S_{g=0}}}{\int {\cal D}[Ac\bar ch]\,e^{-S_{g=0}}}\,,\label{eq:prop}
\eeq
where $S_{g=0}$, the non-interacting part of the action (\ref{eq:CF}), is quadratic in the fields:
\beq
S_{g=0}=\int_x\left\{\frac{1}{4}(\partial_\mu A_\nu^a-\partial_\nu A_\mu^a)^2+\partial_\mu\bar c^a\partial_\mu c^a+ih^a\partial_\mu A_\mu^a+\frac{1}{2}m^2A_\mu^a A_\mu^a\right\}.\label{eq:CF0}
\eeq
Determining the free gluon propagator is thus tantamount to evaluating a Gaussian functional integral. Recall then that, in general,
\beq
\frac{\int {\cal D}\varphi\,\varphi_i(x)\varphi_j(y)\,\exp\left\{-\frac{1}{2}\int_u \int_v\,\varphi_h(u)M_{hk}(u,v)\varphi_k(v)\right\}}{\int {\cal D}\varphi\,\exp\left\{-\frac{1}{2}\int_u \int_v\,\varphi_h(u)M_{hk}(u,v)\varphi_k(v)\right\}}=M^{-1}_{ij}(x,y)\,.
\eeq
So, what we need to do in order to evaluate (\ref{eq:prop}) is, first, to identify the relevant quadratic form $M$, and, then, to find its inverse $M^{-1}$ such that
\beq
\int_z M_{ik}(x,z)M^{-1}_{kj}(z,y)=\delta_{ij}\delta(x-y)\,.\label{eq:inv}
\eeq 
To simplify the discussion, we note that, for a translationally invariant system, such as the one considered here, we have $M_{ij}(x,y)\equiv M_{ij}(x-y)$ and, consequently, $M^{-1}_{ij}(x,y)\equiv M^{-1}_{ij}(x-y)$, so that Eq.~(\ref{eq:inv}) reads
\beq
\int_z M_{ik}(x-z)M^{-1}_{kj}(z-y)=\delta_{ij}\delta(x-y)\,.
\eeq 
This convolution in direct space can be turned into an algebraic relation in Fourier space. Indeed, one finds {\bf [check it]}
\beq
M_{ik}(Q)M^{-1}_{kj}(Q)=\delta_{ij}\,,
\eeq
with
\beq
M_{ij}(Q)\equiv\int_x e^{iQ\cdot x}M_{ij}(x) \quad {\rm and} \quad M^{-1}_{ij}(Q)\equiv\int_x e^{iQ\cdot x}M^{-1}_{ij}(x)\,,
\eeq
the associated Fourier transforms.

Coming back to the Curci-Ferrari model, in order to identify the relevant quadratic form $M$, we note that the ghost fields can be ignored in Eq.~(\ref{eq:CF0}) since they decouple from the evaluation of the free gluon propagator. On the contrary the Nakanishi-Lautrup field $h^a$ should be kept for it couples linearly to $A_\mu^a$ thus leading to a term quadratic in the fields. Using an integration by parts, we can now write $S_{g=0}$ as
\beq
& &S_{g=0}=\frac{1}{2}\int_x\int_y A_\mu^a(x)\delta^{ab}\left[\delta_{\mu\nu}(-\partial^2_x+m^2)+\partial^x_\mu\partial^x_\nu\right]\delta(x-y)A_\nu^b(y)\nonumber\\
& & \hspace{1.0cm}+\,\frac{i}{2}\int_x\int_y h^a(x)\delta^{ab}\partial^x_\nu\delta^x(x-y) A_\nu^b(y)\nonumber\\
& & \hspace{1.0cm}-\frac{i}{2}\int_x\int_y A_\mu^a(x)\delta^{ab}\partial^x_\mu\delta(x-y) h^b(y)\,,
\eeq
which takes the form $\frac{1}{2}\int_x\int_y \varphi_i(x)M_{ij}(x,y)\varphi_j(y)$ provided one gathers the fields $A_\mu^a(x)$ and $h^a(x)$ into a super-field
\beq
\varphi_i(x)=\left\{
\begin{array}{l}
A_\mu^a(x)\,,\quad \mbox{for } i=(\mu,a)\\
h^a(x)\,, \quad \mbox{for } i=a
\end{array}\right.\,.
\eeq
Using a matrix notation, we have
\beq
M(x-y)=\delta^{ab}\left(
\begin{array}{cc}
\delta_{\mu\nu}(-\partial^2_x+m^2)+\partial^x_\mu\partial^x_\nu  & -i\partial^x_\mu\\
i\partial^x_\nu & 0
\end{array}
\right)\delta(x-y)\,,
\eeq
or, in Fourier space,
\beq
M(Q)=\delta^{ab}\left(
\begin{array}{cc}
\delta_{\mu\nu}(Q^2+m^2)-Q_\mu Q_\nu  & -Q_\mu\\
Q_\nu & 0
\end{array}
\right).
\eeq
It is pretty clear that the inverse should have a similar structure
\beq
M^{-1}(Q)=\delta^{ab}\left(
\begin{array}{cc}
\alpha\delta_{\mu\nu}+\beta Q_\mu Q_\nu  & \gamma Q_\mu\\
-\gamma Q_\mu & \delta
\end{array}
\right).\label{eq:Minv}
\eeq
It is actually convenient to re-express the first block of each of these matrices in terms of the longitudinal and transversal projectors
\beq
P^\parallel_{\mu\nu}(Q) & \!\!\!\equiv\!\!\! & \frac{Q_\mu Q_\nu}{Q^2}\,,\\
P^\perp_{\mu\nu}(Q) & \!\!\!\equiv\!\!\! & \delta_{\mu\nu}-\frac{Q_\mu Q_\nu}{Q^2}\,,\
\eeq
which form a complete set of orthogonal projectors {\bf [check it]}
\beq
P^\parallel(Q)+P^\perp(Q) & \!\!\!=\!\!\! & \mathds{1}\,,\\
P^\parallel(Q)^2 & \!\!\!=\!\!\! & P^\parallel(Q)\,,\\
P^\perp(Q)^2 & \!\!\!=\!\!\! & P^\perp(Q)\,,\\ 
P^\perp(Q) P^\parallel(Q) & \!\!\!=\!\!\! & P^\parallel(Q)P^\perp(Q)=0\,,
\eeq
where a matrix notation is implied. We also have {\bf [check it]}
\beq
P^\parallel(Q)\cdot Q=Q\,, \quad P^\perp(Q)\cdot Q=0\,,
\eeq 
where a matrix/vector notation is implied. In terms of these projectors, the matrices to be compared rewrite
\beq
M(Q)=\delta^{ab}\left(
\begin{array}{cc}
(Q^2+m^2)P^\perp_{\mu\nu}(Q)+m^2P^\parallel_{\mu\nu}(Q)  & -Q_\mu\\
Q_\nu & 0
\end{array}
\right)\label{eq:224}
\eeq
and
\beq
M^{-1}(Q)=\delta^{ab}\left(
\begin{array}{cc}
\alpha P^\perp_{\mu\nu}(Q)+\beta P^\parallel_{\mu\nu}(Q)  & \gamma Q_\mu\\
-\gamma Q_\nu & \delta
\end{array}
\right),
\eeq
for some new $\alpha$ and $\beta$ that can be easily related to those in Eq.~(\ref{eq:Minv}). Expressing that $M(Q)$ and $M^{-1}(Q)$ are inverse of each other, one finds the four equations {\bf [check it]}
\beq
\delta_{\mu\nu} & \!\!\!=\!\!\! & \alpha(Q^2+m^2)P^\perp_{\mu\nu}(Q)+(\beta m^2+\gamma Q^2)P^\parallel_{\mu\nu}(Q)\,,\\
0 & \!\!\!=\!\!\! & \beta Q_\mu\,,\\
0 & \!\!\!=\!\!\! & (-m^2\gamma+\delta)Q_\nu\,,\\
1 & \!\!\!=\!\!\! & \gamma Q^2\,,
\eeq
which are solved as
\beq
\alpha=\frac{1}{Q^2+m^2}\,, \quad \beta=0\,, \quad \gamma=\frac{1}{Q^2}\,, \quad \delta=\frac{m^2}{Q^2}\,,
\eeq
and so the propagator in Fourier space reads
\beq
M^{-1}(Q)=\delta^{ab}\left(
\begin{array}{cc}
\frac{P^\perp_{\mu\nu}(Q)}{Q^2+m^2}  & \frac{Q_\mu}{Q^2}\\\\
-\frac{Q_\mu}{Q^2} & \frac{m^2}{Q^2}
\end{array}
\right).
\eeq
Note that this is the propagator in the $A-h$ sector. It contains four blocks corresponding respectively to $\langle AA\rangle$, $\langle Ah\rangle$, $\langle hA\rangle$ and $\langle hh\rangle$. The gluon propagator is to be found in the first block
\beq
\boxed{G_{\mu\nu}^{ab}(Q)=\delta^{ab}\frac{P^\perp_{\mu\nu}(Q)}{Q^2+m^2}}\,\,.\label{eq:231}
\eeq
The last block $\delta^{ab}m^2/Q^2$ corresponds to the Nakanishi-Lautrup propagator while the off-diagonal blocks $\pm \delta^{ab}Q_\mu/Q^2$ are mixed propagators connecting the Nakanishi-Lautrup field to the gluon field. Although non-zero, these propagators do not enter loop calculations since there is no interaction vertex involving the Nakanishi-Lautrup field. In other words, the Nakanishi-Lautrup sector does not receive quantum corrections, see below for further details. This makes actually a lot of sense since the Nakanishi-Lautrup field is just an auxiliary field whose only role is to impose the Landau gauge condition $\smash{\partial_\mu A_\mu^a=0}$.

\subsection{Free Ghost Propagator}
Similarly, the free ghost propagator is defined as
\beq
D^{ab}(x,y)\equiv\frac{\int {\cal D}[Ac\bar ch]\,c^a(x)\bar c^b(y)\,e^{-S_{g=0}}}{\int {\cal D}[Ac\bar ch]\,e^{-S_{g=0}}}\,.
\eeq
Here we need the generalization of Gaussian integration to Grassmanian variables:
\beq
\frac{\int{\cal D}[\theta,\bar\theta]\,\theta_i(x)\bar\theta_j(y)\,\exp\left\{-\int_u\int_v\bar\theta_h(u)M_{hk}(u,v)\theta_k(v)\right\}}{\int{\cal D}[\theta,\bar\theta]\,\exp\left\{-\int_u\int_v\bar\theta_h(u)M_{hk}(u,v)\theta_k(v)\right\}}=M^{-1}_{ij}(x,y)\,.
\eeq
We have this time
\beq
M(x-y)=-\delta^{ab}\partial^2_x\delta(x-y)\,,
\eeq
and thus
\beq
M(Q)=\delta^{ab}Q^2\,,
\eeq
whose inverse gives the free ghost propagator
\beq
\boxed{D^{ab}(Q)=\frac{\delta^{ab}}{Q^2}}\,\,.
\eeq

\subsection{Tree-level Vertices}
Looking now at the interacting part of the action in the ghost sector, we find
\beq
\int_x gf^{abc}\partial_\mu\bar c^a(x) A_\mu^b(x) c^c(x)\,.
\eeq
In Fourier space, this becomes
\beq
-igf^{abc}\int_{P,Q,R}(2\pi)^d\delta(P+Q+R)\,P_\mu \bar c^a(P) A_\mu^b(Q) c^c(R)\,,
\eeq
with $\smash{\int_P\equiv \int d^dP/(2\pi)^d}$. Taking into account the minus sign in front of the action in the Euclidean functional integral, this gives the Feynman rule
\beq
\boxed{igf^{abc}P_\mu}\,\,,\label{eq:240}
\eeq
where $a$, $b$ and $c$ are the colors carried by the antighost, gluon and ghost respectively and $P$ the momentum carried by the antighost.

There are also pure glue vertices that appear in the expansion of $F_{\mu\nu}^aF_{\mu\nu}^a$. We shall write them explicitly in the next lecture. For the time being, we need only to know that the three-gluon vertex is a symmetrized version of the ghost-antighost-gluon vertex, whereas the four-gluon vertex carries no momentum.

\section{Some All-Order Properties}
Let us now use the Feynman rules to derive some properties obeyed by the correlation/vertex functions of the Curci-Ferrari model to all orders.

\subsection{Gluon Propagator}
We have found above that the free gluon propagator is transverse, in the sense that $\smash{Q_\mu G_{\mu\nu}^{ab}(Q)=0}$ and similarly $\smash{G_{\mu\nu}^{ab}(Q)Q_\nu=0}$. This is because of the presence of the transverse projector $P^\perp_{\mu\nu}(Q)$ in Eq.~(\ref{eq:231}). In fact, this could have been anticipated before any explicit calculation. Indeed, because we have restricted to Landau gauge configurations $\smash{\partial_\mu A_\mu^a=0}$, we have
\beq
\partial_\mu G^{ab}_{\mu\nu}(x)=\frac{\int {\cal D}[Ac\bar ch]\,\overbrace{\partial_\mu A^a_\mu(x)}^{\to\,0}A^b_\nu(0)\,e^{-S_{g=0}}}{\int {\cal D}[Ac\bar ch]\,e^{-S_{g=0}}}=0\,,
\eeq
which leads precisely to the transversality condition in Fourier space. 

Let us now observe that this argument extends trivially to the non-interacting case, meaning that the exact gluon propagator
\beq
{\cal G}_{\mu\nu}^{ab}(x,y)\equiv\frac{\int {\cal D}[Ac\bar ch]\,A^a_\mu(x)A^b_\nu(y)\,e^{-S}}{\int {\cal D}[Ac\bar ch]\,e^{-S}}\,,
\eeq
should be transverse as well, or in other words that
\beq
\boxed{{\cal G}_{\mu\nu}^{ab}(Q)=\delta^{ab}P^\perp_{\mu\nu}(Q){\cal G}(Q)}\,\,,\label{eq:243}
\eeq
for some scalar function ${\cal G}(Q)$ to be determined.\footnote{We shall evaluate it at one-loop order in the next lecture.}

\subsection{Ghost Propagator}
Consider now the exact ghost propagator
\beq
{\cal D}^{ab}(x,y)\equiv\frac{\int {\cal D}[Ac\bar ch]\,c^a(x)\bar c^b(y)\,e^{-S}}{\int {\cal D}[Ac\bar ch]\,e^{-S}}\,.
\eeq
As is well known, in Fourier space, it is more convenient to determine its inverse
\beq
{\cal D}^{-1}_{ab}(Q)=Q^2\delta^{ab}+\Sigma^{ab}(Q)\,,
\eeq
where $\smash{\Sigma^{ab}(Q)=\delta^{ab}\Sigma(Q)}$ is known as the self-energy and is given by $-1$ times the sum of one-particle-irreducible (1PI) diagrams attached to one external ghost leg and one external antighost leg. 

\begin{figure}[t]
\begin{center}
\includegraphics[height=0.15\textheight]{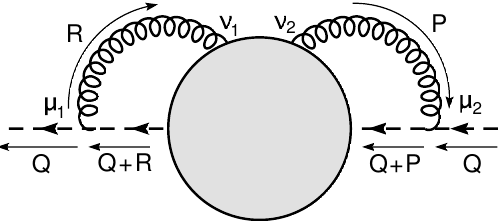}\\
\caption{General structure of a diagram contributing to the ghost self-energy.}\label{fig:Taylor}
\end{center}
\end{figure}

Consider now any such diagram. The antighost leg is attached to a ghost-antighost-gluon vertex which produces a factor $Q_{\mu_1}$, where $\mu_1$ is the Lorentz index carried by the internal gluon line attached to that vertex. On the other hand, the ghost leg is also attached to a ghost-gluon vertex (necessarily different from the previous one). This vertex produces a factor $(Q+P)_{\mu_2}$ where $\mu_2$ is the Lorentz index carried by the internal gluon line (which could coincide with the other one) attached to that vertex and $P$ is the corresponding momentum, so that $Q+P$ is the momentum carried by the antighost attached to that same vertex. But this gluon line also comes with a projector $P^\perp_{\nu_2\mu_2}(P)$. All together, this produces the factor
\beq
P^\perp_{\nu_2\mu_2}(P)(Q+P)_{\mu_2}=P^\perp_{\nu_2\mu_2}(P)Q_{\mu_2}+\underbrace{P^\perp_{\nu_2\mu_2}(P)P_{\mu_2}}_{\to\,0}\,,
\eeq
that is an extra factor of $Q$. Since the ghost self-energy is a scalar, these two factors of $Q$ produce an overall factor of $Q^2$ and lead to the conclusion that $\Sigma(Q^2)/Q^2$ should be well defined in the limit $\smash{Q^2\to 0}$.\footnote{Actually, one needs to use the fact that the gluon propagator is massive to complete the argument.} In other words, the exact ghost propagator can be written as
\beq
\boxed{{\cal D}_{ab}(Q)=\frac{\delta_{ab}}{Q^2}{\cal F}(Q^2)}\,\,,\label{eq:246}
\eeq
where $\smash{{\cal F}(Q^2)=1/(1+\Sigma(Q)/Q^2)}$ is regular as $\smash{Q^2\to 0}$ and is known as the ghost dressing function.

\subsection{Ghost-Antighost-Gluon Vertex}
We can repeat the same argument for any one-particle-irreducible vertex function involving ghost and antighost legs (necessarily equal in number). For each such leg, there will be a factor carrying the corresponding external momentum. 

A direct consequence of this simple observation is that the superficial degree of divergence of any vertex involving ghost and antighost legs is lower than the corresponding function involving only gluon legs. For instance, while the four-gluon vertex will be found to diverge logarithmically, see below, the two-ghost/two-antighost vertex is finite. This is good news, otherwise one would need to modify the original action by a four-ghost interaction in order to absorb the corresponding divergences. The renormalizability of the Curci-Ferrari model will be discussed in more detail below.

A particularly interesting result is obtained in the case of the ghost-antighost-gluon vertex. From the general discussion above, the loop corrections to this vertex vanish when either the ghost or the antighost momentum is taken to zero. But this vertex has also a tree-level contribution as given, up to an overall sign, by the associated Feynman rule (\ref{eq:240}). The latter vanishes only when the antighost momentum is taken to zero,  so one finds
\beq
\lim_{P\to 0}\Gamma^{(3)}_{\bar\psi^aA_\mu^b\psi^c}(P,Q,R)=0\,,
\eeq
but
\beq
\boxed{\lim_{R\to 0}\Gamma^{(3)}_{\bar\psi^aA_\mu^b\psi^c}(P,Q,R)=-igf^{abc}P_\mu}\,\,.\label{eq:Taylor}
\eeq
This result is at the origin of one of the non-renormalization theorems that we shall discuss below.

\section{Generating Functionals}
So far, we looked at the general properties of some of the correlation/vertex functions. A very fruitful approach is to consider all of them at once. This is done by means of generating functionals. Deducing general properties on these functionals, then allows one to deduce properties on all correlation/vertex functions.

\subsection{Definitions}
In particular, one defines
\beq
Z[J,\eta,\bar\eta,R]\equiv\int{\cal D}[Ac\bar ch]\,e^{-S+\int_x (J_\mu^aA_\mu^a+\bar\eta^a c^a+\bar c^a\eta^a+R^aih^a)}\,,\label{eq:Z}
\eeq
the generating functional for $n$-point correlation functions, or, equivalently,
\beq
W[J,\eta,\bar\eta,R]\equiv\ln Z[J,\eta,\bar\eta,R]\,,
\eeq
the generating functional for connected $n$-point correlation functions. Finally, Legendre transforming this functional with respect to the sources, one obtains
\beq
\Gamma[A,c,\bar c,ih^a]\equiv -W[J,\eta,\bar\eta,R]+\int_x(J_\mu^aA_\mu^a+\bar\eta^a c^a+\bar c^a\eta^a+R^aih^a)\,,
\eeq
the generating functional for  one-particle-irreducible (1PI) $n$-point functions or vertex functions $\Gamma^{(n)}$ obtained as iterated functional derivatives of $\Gamma$ with respect to $A$, $c$, $\bar c$ or $ih$. 

Here, the variables $A$, $c$, $\bar c$ and $ih$ are related to the sources $J$, $\eta$, $\bar\eta$ and $R$ by (the signs have to do with the Grassmanian nature of the ghost fields and associated sources)
\beq
A_\mu^a=\frac{\delta W}{\delta J_\mu^a}\,, \quad c^a=\frac{\delta W}{\delta\bar\eta^a}\,, \quad \bar c^a=-\frac{\delta W}{\delta\eta^a}\,, \quad ih^a=\frac{\delta W}{\delta R^a}\,,\label{eq:sources1}
\eeq
and reciprocally
\beq
J_\mu^a=\frac{\delta\Gamma}{\delta A_\mu^a}\,, \quad \eta^a=\frac{\delta\Gamma}{\delta\bar c^a}\,, \quad \bar\eta^a=-\frac{\delta\Gamma}{\delta c^a}\,, \quad R^a=\frac{\delta\Gamma}{\delta ih^a}\,.\label{eq:sources2}
\eeq
The functional $\Gamma$ is also known as the effective action since its tree-level contribution (that is the leading contribution in a loop expansion) coincides with the classical or microscopic theory $S$, while the difference $\Gamma-S$ contains the quantum fluctuations that bring corrections to this microscopic theory. Therefore, deriving the properties of $\Gamma$ allows one to deduce all-order properties of the corresponding vertex functions.

\subsection{Nakanishi-Lautrup Sector}
One first result in that direction is that the quantum fluctuations contained in $\Gamma-S$ do not depend on the Nakanishi-Lautrup field. In other words, the Nakanishi-Lautrup sector of $S$ does not get renormalized. One way to argue this is to shift the Nakanishi-Lautrup field by an arbitrary infinitesimal field, $h^a(x)\to h^a(x)+\varepsilon^a(x)$. Then
\beq
Z[J,\eta,\bar\eta,R] & \!\!\!=\!\!\! & \int{\cal D}[Ac\bar ch]\,e^{-S-\int_x i\varepsilon^a\partial_\mu A_\mu^a+\int_x (J_\mu^aA_\mu^a+\bar\eta^a c^a+\bar c^a\eta^a+R^ai(h^a+\varepsilon^a))}\nonumber\\
& \!\!\!=\!\!\! & \int{\cal D}[Ac\bar ch]\,e^{-S+\int_x (J_\mu^aA_\mu^a+\bar\eta^a c^a+\bar c^a\eta^a+R^aih^a)}\nonumber\\
& & \hspace{1.7cm}\times\,\left[1+i\int_x\varepsilon^a(x)(-\partial_\mu A_\mu^a(x)+R^a(x))\right].
\eeq
Subtracting (\ref{eq:Z}) and using that $\varepsilon^a(x)$ is arbitrary, we arrive at
\beq
0=\int{\cal D}[Ac\bar ch]\,e^{-S+\int_x (J_\mu^aA_\mu^a+\bar\eta^a c^a+\bar c^a\eta^a+R^aih^a)}(-\partial_\mu A_\mu^a(x)+R^a(x))\,.
\eeq
This is nothing but
\beq
\partial_\mu\frac{\delta Z}{\delta J_\mu^a(x)}=R^a(x)Z\,,
\eeq
or, better,
\beq
\partial_\mu\frac{\delta W}{\delta J_\mu^a(x)}=R^a(x)\,.
\eeq
Owing to Eqs.~(\ref{eq:sources1})-(\ref{eq:sources2}), this rewrites
\beq
\boxed{\frac{\delta\Gamma}{\delta ih^a(x)}=\partial_\mu A_\mu^a(x)}\,\,.
\eeq
Since the classical action obeys the same equation, we deduce that $\Gamma-S$ does not depend on $h$, as announced.

\subsection{Back to the Propagator}
The $n$-point vertex functions are directly obtained from the iterated field derivative of the effective action, $\Gamma^{(n)}$. We stress that, for $\smash{n=2}$, this is not the propagator but rather its inverse. To obtain the propagator, one then needs to invert $\Gamma^{(2)}$. 

In particular, to obtain the gluon propagator, we need to invert the matrix formed by $\Gamma^{(2)}_{AA}$, $\Gamma^{(2)}_{Ah}$,  $\Gamma^{(2)}_{hA}$ and $\Gamma^{(2)}_{hh}$. But since the Nakanishi-Lautrup sector is not renormalized, this matrix has a similar structure as the one in Eq.~(\ref{eq:224}):
\beq
\delta^{ab}\left(
\begin{array}{cc}
\Gamma_{\mu\nu}(Q)  & -Q_\mu\\
Q_\nu & 0
\end{array}
\right),
\eeq
where we have used that $\smash{\Gamma^{(2)}_{A_\mu^a A_\nu^b}(Q)\equiv\delta^{ab}\Gamma_{\mu\nu}(Q)}$. Decomposing $\Gamma_{\mu\nu}(Q)$ along longitudinal and transversal projectors:
\beq
\Gamma_{\mu\nu}(Q)=\Gamma_\parallel(Q)P^\parallel_{\mu\nu}(Q)+\Gamma_\perp(Q)P^\perp_{\mu\nu}(Q)\,,
\eeq
one can identically repeat the calculation done in the free case, to find the exact propagator
\beq
\delta^{ab}\left(
\begin{array}{cc}
\frac{P^\perp_{\mu\nu}(Q)}{\Gamma_\perp(Q)}  & \frac{Q_\mu}{Q^2}\\\\
-\frac{Q_\nu}{Q^2} & \frac{\Gamma_\parallel(Q)}{Q^2}
\end{array}
\right).\label{eq:261}
\eeq
The gluon propagator is to be found in the first block and we recover Eq.~(\ref{eq:243}), with
\beq
\boxed{{\cal G}(Q)=\frac{1}{\Gamma_\perp(Q)}}\,\,.\label{eq:162}
\eeq
Similarly, one can show that (mind the signs), setting $\Gamma_{c^a\bar c^b}(Q)\equiv\delta^{ab}\Gamma(Q)$, one finds that the ghost dressing function in Eq.~(\ref{eq:246}) is given by
\beq
\boxed{{\cal F}(Q)=\frac{Q^2}{\Gamma(Q)}}\,\,.\label{eq:163}
\eeq
For latter purpose, and as can be read off from the last block of the matrix (\ref{eq:261}), we note that the Nakanishi-Lautrup propagator is directly connected to $\Gamma_\parallel(Q)$:
\beq
\boxed{\langle h^ah^b\rangle=\delta^{ab}\frac{\Gamma_\parallel(Q)}{Q^2}}\,\,.\label{eq:164}
\eeq
Even though this propagator does not enter loop calculation, this connection to $\Gamma_\parallel(Q)$ will be useful below to derive a second non-renormalization theorem.

\section{Modified BRS Symmetry}
Other important properties of $\Gamma$ stem from the symmetries of the problem. Let us here discuss the case of a modified version of the Becchi-Rouet-Stora-Tyutin (BRST) symmetry obeyed by the Curci-Ferrari model.

\subsection{Faddeev-Popov Case}
The Faddeev-Popov action is well known for its underlying BRST symmetry. One way to derive it is to notice that the ghost term in the action involves a sort of infinitesimal gauge transformation $D_\mu c^a$. It is then natural to consider the following transformation of the gauge field
\beq
\delta A_\mu^a \equiv\bar\omega D_\mu c^a\equiv \bar\omega^a sA_\mu^a\,,
\eeq
with $\bar\omega$ a constant Grassmanian parameter, and to look for similar transformations of the other fields $\smash{\delta\varphi^a=\bar\omega s\varphi^a}$ such that the action is left invariant. One has
\beq
sS_{\rm FP}=\int_x\Big\{\partial_\mu s\bar c^a sA_\mu^a-\bar c^a s^2 A_\mu^a+sih^a\partial_\mu A_\mu^a+ih^a \partial_\mu sA_\mu\Big\}\,,
\eeq
where we have used that the YM term is invariant on its own. Comparing the first and last terms (while using an integration by parts), we observe that we can cancel them by setting
\beq
s\bar c^a\equiv ih^a\,.
\eeq
Similarly, the third term can be cancelled by setting
\beq
sih^a\equiv 0\,.
\eeq
We are left with the second term which requires evaluating $s^2A_\mu^a$. We find
\beq
s^2A_\mu^a & \!\!\!=\!\!\!\! & sD_\mu c^a\nonumber\\
& \!\!\!=\!\!\! & D_\mu sc^a+gf^{abc}sA_\mu^b c^c\nonumber\\
& \!\!\!=\!\!\! & D_\mu sc^a+gf^{abc}D_\mu c^b c^c\nonumber\\
& \!\!\!=\!\!\! & D_\mu sc^a+D_\mu\left(\frac{1}{2}gf^{abc}c^b c^c\right),
\eeq
which we can cancel provided we define
\beq
sc^a\equiv-\frac{1}{2}gf^{abc}c^bc^c\,.
\eeq
All in all, we have shown that the Faddeev-Popov action is invariant under the transformation
\beq
sA_\mu^a=D_\mu c^a\,, \quad sc^a=-\frac{g}{2}f^{abc}c^bc^c\,, \quad s\bar c^a=ih^a\,, \quad sih^a=0\,.
\eeq
This is known as BRST symmetry. By construction $\smash{s^2A_\mu^a=0}$. It is also trivial to check that $\smash{s^2\bar c^a=0}$ and $\smash{s^2ih^a=0}$ and, less trivial, but also possible to check {\bf [do it]} that $\smash{s^2c^a=0}$. This implies that $s^2=0$ over the whole field space. The BRST transformation is said to be nilpotent.

\subsection{Curci-Ferrari Model}
What about the Curci-Ferrari model? Let us first analyze how the mass term transforms under $s$. We have
\beq
s\int_x\frac{1}{2}m^2A_\mu^aA_\mu^a=m^2\int_x A_\mu^a D_\mu c^a=m^2\int_x A_\mu^a \partial_\mu c^a=-m^2\int_x \partial_\mu A_\mu^a  c^a\,,
\eeq
which can be cancelled if instead of transforming $ih^a$ as above, we define a new transformation $\hat s$ as follows
\beq
\boxed{\hat sA_\mu^a=D_\mu c^a\,, \quad \hat sc^a=-\frac{g}{2}f^{abc}c^bc^c\,, \quad \hat s\bar c^a=ih^a\,, \quad \hat sih^a=m^2c^a}\,\,.
\eeq
This is known as the mBRS symmetry ($m$ for modified or massive). We have again $\hat s^2A_\mu^a=0$, $\hat s^2c^a=0$. However, $\hat s^2\bar c^a=m^2c^a$ and $\hat s^2ih^a=m^2\hat sc^a$. So, $\hat s$ is nilpotent only in the sector $A-c$.

\subsection{Constraint on the Dressing Function}
There is a very simple constraint of the mBRS symmetry. Let us consider the correlator $\langle ih^a\bar c^b \rangle$ (which is actually equal to $0$ because its ghost number is non-zero). The mBRST symmetry implies that this correlator rewrites (we use $\bar\omega^2=0$)
\beq
\langle ih^a\bar c^b\rangle & = & \langle (ih^a+\bar\omega \hat sih^a)(\bar c^b+\bar\omega \hat s\bar c^b)\rangle\nonumber\\
& = & \langle ih^a\bar c^b\rangle+\bar\omega(m^2\langle c^a\bar c^b\rangle-\langle h^ah^b\rangle)\,,
\eeq
and thus
\beq
m^2\langle c^a\bar c^b\rangle=\langle h^ah^b\rangle\,.
\eeq
But using (\ref{eq:246}) and (\ref{eq:261}) while switching to Fourier space, we find
\beq
m^2\frac{{\cal F}(Q)}{Q^2}=\frac{\Gamma^\parallel(Q)}{Q^2}
\eeq
or
\beq
\boxed{\Gamma^\parallel(Q){\cal F}^{-1}(Q)=m^2}\,\,.\label{eq:275}
\eeq

\section{Renormalization}
In four dimensions, the dimension of the coupling vanishes which tells us that the UV divergences are limited to a finite number of vertex functions and can thus be absorbed by adding to the original action local operators of dimension less or equal to $4$. The question is whether the so-modified action amounts to a simple rescaling of the fields and parameters of the original action
\beq
A\to Z_A^{1/2}A\,, \quad c\to Z_c^{1/2}c\,, \quad g\to Z_g g\,, \quad m^2\to Z_{m^2}m^2\,.
\eeq
The answer to this question is positive but the proof goes beyond the scope of these lectures.\footnote{The Curci-Ferrari model should not be mistaken with the non-Abelian version of Proca theory
$$S_{\rm Proca}=\int_x \left\{\frac{1}{4}F_{\mu\nu}^aF_{\mu\nu}^a+\frac{1}{2}m^2A_\mu^a A_\mu^a\right\}\neq S_{\rm CF}\,.\nonumber$$
The latter theory is non-renormalizable while the CF model is renormalizable. Let us also mention that the Proca theory is an intentional modification of a fundamental theory while the CF model is an attempt at modifying the gauge-fixed action, which needs to be modified anyway due to the presence of Gribov copies. Finally, the Proca theory breaks gauge invariance while in the CF model, gauge invariance is already broken by the gauge fixing via the FP terms.} We give however a substantial account of this proof in the Appendix.

In the remainder of this first lecture, we limit ourselves to deduce two non-renormalization theorems that allow one to relate the divergences of the four renormalization factors given above.

\subsection{Coupling Renormalization}
Consider first the ghost-antighost-gluon vertex. We have seen that it obeys the strong contraint (\ref{eq:Taylor}). Upon renormalization, this equation becomes
\beq
\lim_{R\to 0}Z_c^{-1}Z_A^{-1/2}\Gamma^{(3)}_{\bar\psi^aA_\mu^b\psi^c}(P,Q,R)=-iZ_g gf^{abc}P_\mu\,,
\eeq
where we stress that to get the renormalization factors in the right positions, one needs to pay attention to the fact that the vertex functions are amputated of their external propagators. Putting all renormalization factors on the same side, the equation becomes
\beq
\lim_{R\to 0}\Gamma^{(3)}_{\bar\psi^aA_\mu^b\psi^c}(P,Q,R)=-iZ_g Z_A^{1/2}Z_c gf^{abc}P_\mu\,.
\eeq
Since both the renormalized vertex and the renormalized coupling are finite, one deduces that
\beq
\boxed{Z_{g}Z_A^{1/2}Z_c \mbox{ is finite}}\,\,.
\eeq

\subsection{Mass Renormalization}
A similar argument shows that the renormalized version of (\ref{eq:275}) is
\beq
Z_A^{-1}Z_c^{-1}\Gamma^\parallel(Q){\cal F}^{-1}(Q)=Z_{m^2}m^2\,,
\eeq
which rewrites
\beq
\Gamma^\parallel(Q){\cal F}^{-1}(Q)=Z_{m^2}Z_A Z_c m^2\,,
\eeq
and from which it follows that
\beq
\boxed{Z_{m^2}Z_AZ_c \mbox{ is finite}}\,\,.
\eeq

\chapter{One-loop Propagators}

In this lecture, we evaluate the two-point Curci-Ferrari correlators at one-loop order in view of comparing them to lattice data for the Landau gauge YM correlators. As we have summarized in Eqs.~(\ref{eq:162})-(\ref{eq:163}), we only need to determine the appropriate second derivatives of the effective action which we have denoted $\Gamma(Q)$ and $\Gamma_{\mu\nu}(Q)$. For the latter, we only need its transversal component $\Gamma_\perp(Q)$. Moreover, we have already evaluated these functions at tree-level
\beq
\Gamma(Q)=Q^2+\dots\,, \quad \Gamma_\perp(Q)=Q^2+m^2+\dots
\eeq 
We now would like to determine the one-loop corrections as given by the one-loop self-energies.

\section{Feynman Rules}
We have already seen in the previous lecture that the free ghost and gluon propagators are given respectively by
\beq
\frac{\delta^{ab}}{Q^2} \quad {\rm and} \quad \delta^{ab}\frac{P^\perp_{\mu\nu}(Q)}{Q^2+m^2}\,,
\eeq
whereas the tree-level ghost-antighost-gluon vertex reads
\beq
igP_\mu f^{abc}\,,
\eeq
where $a$ and $P$ are associated to the antighost leg of the vertex, $b$ and $\mu$ to the gluon leg and $c$ to the ghost leg.

The YM term of the action leads in addition to three- and four-gluon tree-level vertices. The former takes the form of a symmetrized version of the ghost-antighost-gluon vertex:
\beq
\frac{ig}{3!}f^{abc} \Big[(P-R)_\nu\delta_{\mu\rho}+(R-Q)_\mu\delta_{\rho\nu}+(Q-P)_\rho\delta_{\nu\mu}\Big]\,,\label{eq:v3}
\eeq
where $(a,P,\mu)$, $(b,Q,\nu)$ and $(c,R,\rho)$ are associated to each leg. As for the four-gluon vertex, it reads
\beq
& & -\frac{g^2}{4!}\Bigg[f^{abe}f^{cde}(\delta_{\mu\rho}\delta_{\nu\sigma}-\delta_{\mu\sigma}\delta_{\nu\rho})\nonumber\\
& & \hspace{1.5cm}+\,f^{ace}f^{dbe}(\delta_{\mu\sigma}\delta_{\rho\nu}-\delta_{\mu\nu}\delta_{\rho\sigma})\nonumber\\
& & \hspace{2.5cm}+\,f^{ade}f^{bce}(\delta_{\mu\nu}\delta_{\sigma\rho}-\delta_{\mu\rho}\delta_{\sigma\nu})\Bigg]\,.
\eeq
Let us also recall that each ghost loop in a diagram requires an additional factor of $-1$.

\begin{figure}[t]
\begin{center}
\includegraphics[height=0.35\textheight]{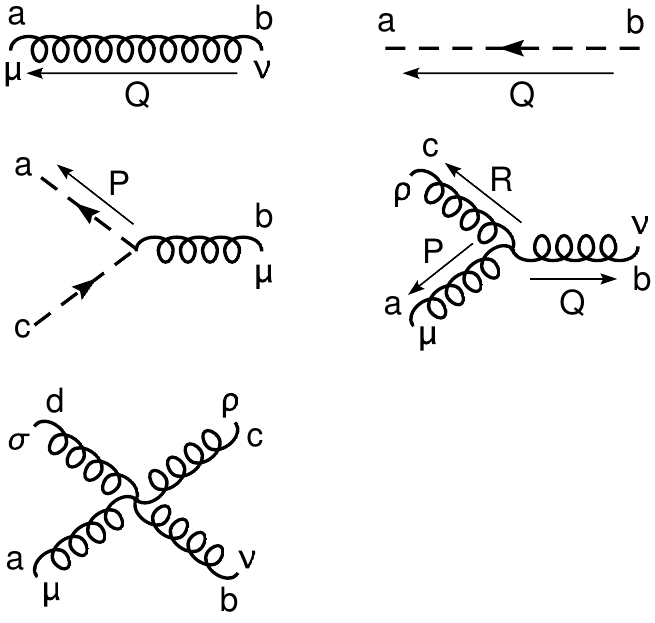}\\
\caption{Diagrammatic representation of the Feynman rules.}\label{fig:rules}
\end{center}
\end{figure}

The set of Feynman rules are represented in Fig.~\ref{fig:rules}.

\section{Feynman Diagrams}

\begin{figure}[t]
\begin{center}
\includegraphics[height=0.25\textheight]{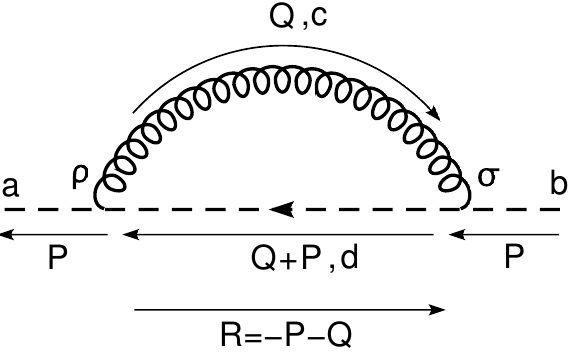}\\
\caption{One-loop diagram contributing to the ghost self-energy.}\label{fig:ghost_self}
\end{center}
\end{figure}

\subsection{Ghost Self-energy}
There is only one diagram contributing to the ghost self-energy at one-loop order, see Fig.~\ref{fig:ghost_self}. It writes\footnote{In our Euclidean convention, the self-energy is given by minus one times the sum of diagrams.}
\beq
\Sigma^{ab}(P) & \!\!\!=\!\!\! & -\frac{1}{2!}\times(ig)^2\times 2\times f^{acd}f^{dcb}\nonumber\\
&  & \times\int_Q  P_\rho P^\perp_{\rho\sigma}(Q)(-R_\sigma)\frac{1}{Q^2+m^2}\frac{1}{R^2}\,,
\eeq
where, for convenience, we have introduced $R$ such that $\smash{P+Q+R=0}$. The color trace is easily done using $\smash{f^{acd}f^{bcd}=N\delta^{ab}}$ and we arrive at $\Sigma^{ab}(P)=\delta^{ab}\Sigma(P)$, with
\beq
\frac{\Sigma(P)}{g^2N}=\int_Q \smash{(P\cdot P^\perp(Q)\cdot R)}\,G(Q)G_0(R)\,,
\eeq
where we have introduced a matrix/vector notation and defined
\beq
G(Q)\equiv\frac{1}{Q^2+m^2} \quad \mbox{and} \quad G_0(R)\equiv\frac{1}{R^2}\,.
\eeq
We can slightly massage the above expression by using
\beq
P^\perp(Q)\cdot P= P^\perp(Q)\cdot (-Q-R)=-P^\perp(Q)\cdot R\,.\label{eq:38}
\eeq 
This, together with
\beq
R\cdot P^\perp(Q)\cdot R=R^2-\frac{(Q\cdot R)^2}{Q^2}=(Q^2R^2-(Q\cdot R)^2)G_0(Q)\,,\label{eq:39}
\eeq
leads to
\beq
\frac{\Sigma(P)}{g^2N}=-\int_Q (Q^2R^2-(Q\cdot R)^2)\,G_0(Q)G(Q)G_0(R)\,.
\eeq
where we note the appearance of two propagators carrying the same momentum. This will be dealt with below.

The combination $Q^2R^2-(Q\cdot R)^2$ is full of simplifying magic. Indeed, using Eq.~(\ref{eq:38}) twice in the LHS of (\ref{eq:39}), we see that this combination is invariant upon exchanging $R$ by $P$. From this, it actually follows that it is invariant under any permutation of the triplet $(P,Q,R)$. We shall denote it by $\sigma(P,Q,R)$ in what follows, or even $\sigma$ for short. At the end of the day, we can write
\beq
\boxed{\frac{\Sigma(P)}{g^2N}=-\int_Q \sigma\,G_0(Q)G(Q)G_0(R)}\,\,.\label{eq:311}
\eeq

\begin{figure}[t]
\begin{center}
\includegraphics[height=0.5\textheight]{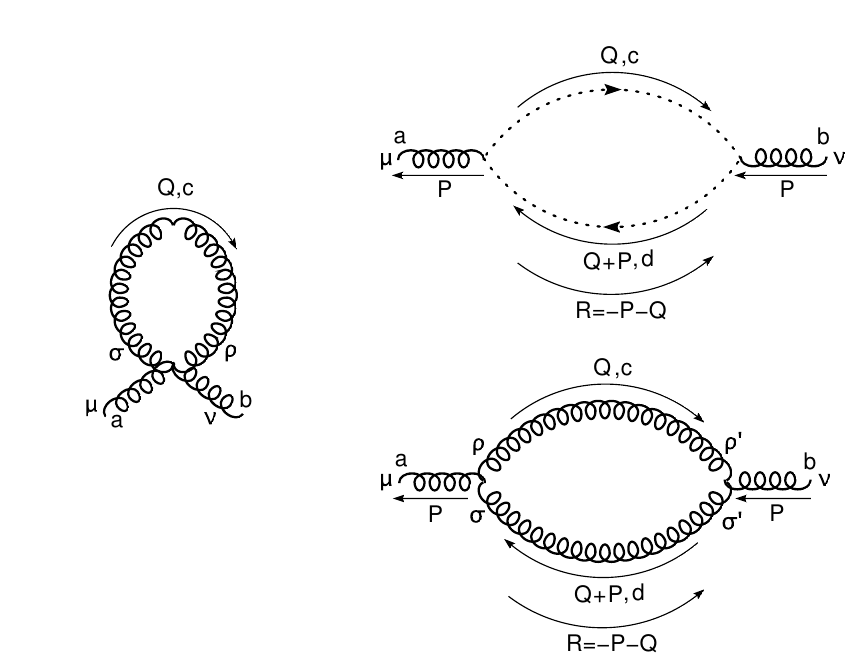}\\
\caption{One-loop diagrams contributing to the gluon self-energy.}\label{fig:gluon_self}
\end{center}
\end{figure}

\subsection{Gluon Self-energy}
The gluon self-energy is made of three diagrams at one-loop order, see Fig.~\ref{fig:gluon_self}. It contains both a longitudinal and a transversal part:
\beq
\Pi^{ab}_{\mu\nu}(P)=\Pi^{ab}_\parallel(P)P^\parallel_{\mu\nu}(P)+\Pi^{ab}_\perp(P)P^\perp_{\mu\nu}(P)\,,
\eeq
but as we have already discussed, we only need its transversal component $\Pi^{ab}_\perp(P)$, which can be obtained via contraction with the transversal projector $P^\perp_{\mu\nu}(P)$ {\bf [check it]}
\beq
P^\perp_{\nu\mu}(P)\Pi^{ab}_{\mu\nu}(P)=(d-1)\Pi^{ab}_\perp(P)\,.
\eeq
So we only need to look at the transversal part of each of the diagrams.

\subsubsection{a. The tadpole diagram}
For instance, the tadpole diagram reads
\beq
(d-1)\Pi^{ab}_{tad,\perp}(P) & \!\!\!=\!\!\! & -\left(-\frac{g^2}{4!}\right)\times 4\times 3\nonumber\\
& & \times\int_Q \Bigg[f^{abe}f^{cce}(\delta_{\mu\rho}\delta_{\nu\sigma}-\delta_{\mu\sigma}\delta_{\nu\rho})\nonumber\\
& & \hspace{1.5cm}+\,f^{ace}f^{cbe}(\delta_{\mu\sigma}\delta_{\rho\nu}-\delta_{\mu\nu}\delta_{\rho\sigma})\nonumber\\
& & \hspace{2.0cm}+\,f^{ace}f^{bce}(\delta_{\mu\nu}\delta_{\sigma\rho}-\delta_{\mu\rho}\delta_{\sigma\nu})\Bigg]\nonumber\\
& & \hspace{2.0cm} \times\,P^\perp_{\mu\nu}(P)P^\perp_{\rho\sigma}(Q)G(Q)\,.
\eeq
The first line vanishes due to $f^{cce}=0$ and it is easily checked that the other two lines contribute identically. The color trace can again be easily computed using $f^{acd}f^{bcd}=N\delta^{ab}$, so we arrive at $\Pi^{ab}_{tad,\perp}(P)=\delta^{ab}\Pi_{tad,\perp}(P)$, with
\beq
(d-1)\frac{\Pi_{tad,\perp}(P)}{g^2N} & \!\!\!=\!\!\! & \int_Q (\delta_{\mu\nu}\delta_{\sigma\rho}-\delta_{\mu\rho}\delta_{\sigma\nu})P^\perp_{\mu\nu}(P)P^\perp_{\rho\sigma}(Q)G(Q)\nonumber\\
& \!\!\!=\!\!\! & \int_Q \Big[{\rm tr}\,P^\perp(P) {\rm tr}\,P^\perp(Q)-{\rm tr}\,(P^\perp(P)P^\perp(Q))\Big]G(Q)\,.\nonumber\\
\eeq
Using
\beq
{\rm tr}\,P^\perp(Q)=d-1\,, \quad {\rm and} \quad {\rm tr}\,(P^\perp(P)P^\perp(Q))=d-2+\frac{(P\cdot Q)^2}{P^2Q^2}\,,\label{eq:314}
\eeq
we arrive eventually at
\beq
\boxed{(d-1)\frac{\Pi_{tad,\perp}(P)}{g^2N}=\int_Q \left[d^2-3d+3-\frac{(P\cdot Q)^2}{P^2Q^2}\right]G(Q)}\,\,.
\eeq

\subsubsection{b. The ghost loop}
The ghost loop reads
\beq
(d-1)\Pi^{ab}_{gh,\perp}(P) & = & -(-1)\times\frac{1}{2!}\times(ig)^2\times 2\times f^{cad}f^{dbc}\nonumber\\
& & \times\int_Q Q_\mu P^\perp_{\mu\nu}(P)(-R_\nu)\,G_0(Q)G_0(R)\,,
\eeq
where the extra factor of $-1$ comes from the ghost loop. The color trace is again easily done using $\smash{f^{acd}f^{bcd}=N\delta^{ab}}$ and we arrive at $\Pi^{ab}_{gh,\perp}(P)=\delta^{ab}\Pi_{gh,\perp}(P)$. The Lorentz contractions can be done using similar tricks as above and we find
\beq
\boxed{(d-1)\frac{\Pi_{gh,\perp}(P)}{g^2N}=\frac{1}{P^2}\int_Q \sigma\,G_0(Q)G_0(R)}\,\,.
\eeq

\subsubsection{c. The gluon loop}
The gluon loop is scarier due to the presence of two three-gluon vertices with six terms each, leading potentially to 36 terms to handle. However, some anticipation allows one to make like simpler. 

The trick is to use momentum conservation at each vertex together with the fact that each gluon leg of the vertex is contracted with the corresponding transversal projector (either because each internal gluon line carries such a projector, or because we are projecting transversally with respect to the external momentum). If we take for instance the first two terms the three gluon vertex (\ref{eq:v3}), they will actually appear as
\beq
P^\perp_{\sigma\nu}(Q)(P-R)_\nu\delta_{\mu\rho}=P^\perp_{\sigma\nu}(Q)(P+P+Q)_\nu\delta_{\mu\rho}=2P^\perp_{\sigma\nu}(Q)P_\nu\delta_{\mu\rho}
\eeq
or
\beq
P^\perp_{\sigma\nu}(Q)(P-R)_\nu\delta_{\mu\rho}=P^\perp_{\sigma\nu}(Q)(-Q-R-R)_\nu\delta_{\mu\rho}=-2P^\perp_{\sigma\nu}(Q)R_\nu\delta_{\mu\rho}\,.
\eeq
This means that one can in practice forget one of the momenta appearing in such differences, provided that the vertex is multiplied by a factor of $2$. Notice also that the two vertices of the gluon loop contribute the same up to changing $a$ to $b$, $\mu$ to $\nu$, $(\rho,\sigma)$ to $(\rho'\sigma')$, and an overall sign due the fact that the outgoing momenta in one vertex are opposite to those in the other vertex. With this in mind, we find
\beq
& & (d-1)\Pi^{ab}_{gl,\perp}(P)=-\frac{1}{2}\times\left(\frac{ig}{3!}\right)^2\times6\times 3\times 2\times f^{acd}f^{bcd}(-1)\times2\times 2\nonumber\\
&  & \hspace{0.5cm}\times\int_Q (R_\mu\delta_{\rho\sigma}+Q_\sigma\delta_{\mu\rho}+P_\rho\delta_{\sigma\mu})(R_\nu\delta_{\rho'\sigma'}+Q_{\sigma'}\delta_{\nu\rho'}+P_{\rho'}\delta_{\sigma'\nu})\nonumber\\
& & \hspace{1.5cm}\times\,P^\perp_{\mu\nu}(P)P^\perp_{\rho\rho'}(Q)P^\perp_{\sigma\sigma'}(R)G(Q)G(R)\,,
\eeq
where we recall that $\smash{P+Q+R=0}$ and we have chosen to keep the expression as symmetric as possible in the exchange of these three momenta. After performing the color trace, we arrive at $\Pi^{ab}_{gl,\perp}(P)=\delta^{ab}\Pi_{gl,\perp}(P)$, with
\beq
& & (d-1)\frac{\Pi_{gl,\perp}(P)}{g^2N}\nonumber\\
& & \hspace{0.5cm}\,=-2\int_Q (R_\mu\delta_{\rho\sigma}+Q_\sigma\delta_{\mu\rho}+P_\rho\delta_{\sigma\mu})(R_\nu\delta_{\rho'\sigma'}+Q_{\sigma'}\delta_{\nu\rho'}+P_{\rho'}\delta_{\sigma'\nu})\nonumber\\
& & \hspace{4.5cm}\times\,P^\perp_{\mu\nu}(P)P^\perp_{\rho\rho'}(Q)P^\perp_{\sigma\sigma'}(R)G(Q)G(R)\,.\label{eq:323}
\eeq
We now would like to evaluate the Lorentz contraction
\beq
X & \!\!\!\equiv\!\!\! & (R_\mu\delta_{\rho\sigma}+Q_\sigma\delta_{\mu\rho}+P_\rho\delta_{\sigma\mu})(R_\nu\delta_{\rho'\sigma'}+Q_{\sigma'}\delta_{\nu\rho'}+P_{\rho'}\delta_{\sigma'\nu})\nonumber\\
& &\hspace{2.0cm} \times\,P^\perp_{\mu\nu}(P)P^\perp_{\rho\rho'}(Q)P^\perp_{\sigma\sigma'}(R)\,.
\eeq
We denote each of the $9$ resulting terms as $X_{ij}$ with $i$ and $j$ representing the rank of the term chosen in each bracket. First if we take the first term of each bracket, we find
\beq
X_{11}=R\cdot P^\perp(P)\cdot R\,{\rm tr}(P^\perp(Q)P^\perp(R))\,.
\eeq
It is easily checked {\bf [do it]} that $X_{22}$ and $X_{33}$ are obtained from $X_{11}$ by cyclically permuting $(P,Q,R)$. Then
\beq
\sum_iX_{ii} & \!\!\!=\!\!\! & Q\cdot P^\perp(R)\cdot Q\,{\rm tr}(P^\perp(P)P^\perp(Q))+\mbox{cyclic permutations}\nonumber\\
& \!\!\!=\!\!\! & \sigma\,\left[\frac{d-2}{R^2}+\frac{(P\cdot Q)^2}{P^2Q^2R^2}\right]+\mbox{cyclic permutations}\,,
\eeq
where we have used Eqs.~(\ref{eq:39}) and (\ref{eq:314}). Now comes the magic. Since $\sigma$ is invariant under any permutation of $(P,Q,R)$, the sum over cyclic permutations rewrites
\beq
\sum_iX_{ii}=\sigma\left[(d-2)\left(\frac{1}{P^2}+\frac{1}{Q^2}+\frac{1}{R^2}\right)+\frac{(P\cdot Q)^2+(Q\cdot R)^2+(R\cdot P)^2}{P^2Q^2R^2}\right].\label{eq:327}\nonumber\\
\eeq
which is nice but still involves some nasty scalar products.

But let's press on and consider now the terms $X_{ij\neq i}$. For instance
\beq
X_{12}=R\cdot P^\perp(P)P^\perp(Q)P^\perp(R)\cdot Q\,,
\eeq
which we conveniently rewrite as 
\beq
X_{12}=-Q\cdot P^\perp(P)P^\perp(Q)P^\perp(R)\cdot Q\,,
\eeq
using Eq.~(\ref{eq:38}). It is also easily seen that $\smash{X_{ij}=X_{ji}}$ and that $X_{23}$ and $X_{31}$ are obtained from $X_{12}$ by cyclically permuting $(P,Q,R)$. All in all, we arrive at
\beq
\sum_{i\neq j}X_{ij}=-\,2R\cdot P^\perp(Q)P^\perp(R)P^\perp(P)\cdot R+\,\mbox{cyclic permutations}\,.
\eeq
We now write
\beq
R\cdot P^\perp(Q)P^\perp(R)P^\perp(P)\cdot R & = & \left[R-\frac{R\cdot Q}{Q^2}Q\right]\cdot P^\perp(R)\cdot\left[R-P\frac{P\cdot R}{P^2}\right]\nonumber\\
& = & \frac{(P\cdot R)(R\cdot Q)}{P^2Q^2}Q\cdot P^\perp(R)\cdot P\nonumber\\
& = & -\frac{(P\cdot R)(R\cdot Q)}{P^2Q^2}Q\cdot P^\perp(R)\cdot Q\nonumber\\
& = & -\frac{(P\cdot R)(R\cdot Q)}{P^2Q^2R^2}(Q^2R^2-(Q\cdot R)^2)\nonumber\\
& = & -\frac{(P\cdot R)(R\cdot Q)}{P^2Q^2R^2}\sigma\,,
\eeq
where we notice again the appearance of the magical $\sigma$. We then find
\beq
\sum_{i\neq j}X_{ij}=\sigma\frac{2(P\cdot R)(R\cdot Q)+2(Q\cdot P)(P\cdot R)+2(R\cdot Q)(Q\cdot P)}{P^2Q^2R^2}\,.
\eeq
Combining this with Eq.~(\ref{eq:327}), we recognize
\beq
& & (P\cdot Q)^2+(Q\cdot R)^2+(R\cdot P)^2\nonumber\\
& & \hspace{0.5cm}+\,2(P\cdot R)(R\cdot Q)+2(Q\cdot P)(P\cdot R)+2(R\cdot Q)(Q\cdot P)\nonumber\\
& & \hspace{0.5cm}=\,(P\cdot Q+Q\cdot R+R\cdot P)^2\nonumber\\
& & \hspace{0.5cm}=\,\left(\frac{(P+Q+R)^2-P^2-Q^2-R^2}{2}\right)^2=\frac{(P^2+Q^2+R^2)^2}{4}\,,\nonumber\\
\eeq
where we have used $\smash{P+Q+R=0}$ in the last step. We then get
\beq
\sum_{ij}X_{ij} & = & \sigma\left[(d-2)\left(\frac{1}{P^2}+\frac{1}{Q^2}+\frac{1}{R^2}\right)+\frac{(P^2+Q^2+R^2)^2}{4P^2Q^2R^2}\right]\nonumber\\
& = & \sigma\left[\left(d-\frac{3}{2}\right)\left(\frac{1}{P^2}+\frac{1}{Q^2}+\frac{1}{R^2}\right)+\frac{P^4+Q^4+R^4}{4P^2Q^2R^2}\right],
\eeq
where the nasty scalar products are now gone! (except those within $\sigma$ but they will be taken care of in the next section). Going back to Eq.~(\ref{eq:323}), we find that
\beq
& & \Bigg|\Bigg|(d-1)\frac{\Pi_{gl,\perp}(P)}{g^2N}\nonumber\\
& & \hspace{1.0cm}=\,-\frac{2}{P^2}\left(d-\frac{3}{2}\right)\int_Q \sigma\,G(Q)G(R)\nonumber\\
& & \hspace{1.5cm}-\,4\left(d-\frac{3}{2}\right)\int_Q \sigma\,G_0(Q)G(Q)G(R)\nonumber\\
& & \hspace{1.5cm}-\,\frac{P^2}{2}\int_Q \sigma\,G_0(Q)G(Q)G_0(R)G(R)\nonumber\\
& & \hspace{1.5cm}-\,\frac{1}{P^2}\int_Q \sigma\,G_0(Q)G(Q)R^2G(R)\,.\label{eq:335}
\eeq

\section{Reduction to Master Integrals}
The next step is to reduce the above integrals to a limited number of simpler (master) integrals.

\subsection{Reduction Tricks}
Sometimes, one finds two propagators carrying the same momentum but different masses. One can handle this situation using
\beq
G_0(Q)G(Q) & \!\!\!=\!\!\! & \frac{1}{Q^2(Q^2+m^2)}=\frac{1}{m^2}\left[\frac{1}{Q^2}-\frac{1}{Q^2+m^2}\right]\nonumber\\
& \!\!\!=\!\!\! & \frac{1}{m^2}\Big[G_0(Q)-G(Q)\Big].\label{eq:336}
\eeq
A similar trick allows one to eliminate unwanted powers of $Q^2$ in the numerator:
\beq
Q^2G(Q) & \!\!\!=\!\!\! & \frac{Q^2}{Q^2+m^2}=\frac{Q^2+m^2-m^2}{Q^2+m^2}\nonumber\\
 & \!\!\!=\!\!\! & 1-\frac{m^2}{Q^2+m^2}=1-m^2G(Q)\,.\label{eq:337}
\eeq
When the angular dependence is only in the numerators, one can do the following replacements under the integrals:
\beq
Q_\mu Q_\nu & \!\!\!\to\!\!\! & Q^2\frac{\delta_{\mu\nu}}{d}\,,\label{eq:338}\\
Q_\mu Q_\nu Q_\rho Q_\sigma & \!\!\!\to\!\!\! & (Q^2)^2\frac{\delta_{\mu\nu}\delta_{\rho\sigma}+\delta_{\mu\rho}\delta_{\nu\sigma}+\delta_{\mu\sigma}\delta_{\nu\rho}}{d(d+2)}\,,\\
& \!\!\!\dots\!\!\! & \nonumber
\eeq
When $P+Q+R=0$, scalar products in the numerator can be handled using
\beq
Q\cdot R=\frac{1}{2}\Big[P^2-Q^2-R^2\Big],\label{eq:340}
\eeq
and the resulting unwanted powers of squared momenta can be eliminated using Eq.~(\ref{eq:337}).

\subsection{Ghost Self-energy}
Let us apply the reduction tricks to the ghost self-energy (\ref{eq:311}). Using Eq.~(\ref{eq:336}), we find
\beq
\frac{\Sigma(P)}{g^2N}=\frac{1}{m^2}\int_Q \sigma\,\Big[G(Q)G_0(R)-(m\to 0)\Big].
\eeq
We now need to handle $\sigma G(Q)G_0(R)=(Q^2R^2-(Q\cdot R)^2)G(Q)G(R)$. Using Eq.~(\ref{eq:337}), we first write
\beq
Q^2R^2G(Q)G_0(R)=Q^2G(Q)=1-m^2G(Q)\,.\label{eq:342}
\eeq
Using Eq.~(\ref{eq:340}) and again Eq.~(\ref{eq:337}), we then write
\beq
& & (Q\cdot R)G(Q)G(R)\nonumber\\ 
& & \hspace{1.0cm}=\,\frac{1}{2}\Big[P^2-Q^2-R^2\Big]G(Q)G_0(R)\nonumber\\
& & \hspace{1.0cm}=\,\frac{1}{2}\Big[P^2+m^2-(Q^2+m^2)-R^2\Big]G(Q)G_0(R)\nonumber\\
& & \hspace{1.0cm}=\,\frac{P^2+m^2}{2}G(Q)G_0(R)-\frac{1}{2}G_0(R)-\frac{1}{2}G(Q)\,,
\eeq
and then
\beq
& & (Q\cdot R)^2G(Q)G(R)\nonumber\\ 
& & \hspace{1.0cm}=\,\frac{P^2+m^2}{2}\left[\frac{P^2+m^2}{2}G(Q)G_0(R)-\frac{1}{2}G_0(R)-\frac{1}{2}G(Q)\right]\nonumber\\
& & \hspace{1.0cm}-\,\frac{Q\cdot R}{2}G_0(R)-\frac{Q\cdot R}{2}G(Q)\,.
\eeq
Combining this with Eq.~(\ref{eq:342}), we find
\beq
& & \sigma\,G(Q)G_0(R)\nonumber\\
& & \hspace{1.0cm}=\,1+\left[\frac{Q\cdot R}{2}+\frac{P^2+m^2}{4}\right]G_0(R)+\left[\frac{Q\cdot R}{2}+\frac{P^2-3m^2}{4}\right]G(Q)\nonumber\\
& & \hspace{1.0cm}-\,\frac{(P^2+m^2)^2}{4}\,G(Q)G_0(R)\,.
\eeq
Using $\smash{P+Q+R=0}$, this rewrites
\beq
& & \sigma\,G(Q)G_0(R)\nonumber\\
& & \hspace{1.0cm}=\,\left[-\frac{P\cdot R}{2}+\frac{P^2+m^2}{4}\right]G_0(R)+\left[-\frac{P\cdot Q}{2}+\frac{P^2-m^2}{4}\right]G(Q)\nonumber\\
& & \hspace{1.0cm}-\,\frac{(P^2+m^2)^2}{4}\,G(Q)G_0(R)\,.\label{eq:346}
\eeq
The first term vanishes upon integration in dimensional regularization. In the second term the same is true for the first term in the bracket due to angular integration. Eventually one arrives at
\beq
\boxed{\frac{\Sigma(P)}{g^2N}=\frac{(P^2)^2}{4m^2}I_{00}(P)-\frac{(P^2+m^2)^2}{4m^2}I_{m0}(P)+\frac{P^2-m^2}{4m^2}J_m}\,\,,\label{eq:248}
\eeq
where
\beq
J_m & \!\!\!\equiv\!\!\! & \int_Q G_m(Q)\,,\\
I_{m_1m_2}(P) & \!\!\!\equiv\!\!\! & \int_Q G_{m_1}(Q)G_{m_2}(R)=I_{m_2m_1}(P)\,,
\eeq
are well known integrals with admit explicit expressions {\bf [homework, and App. B for the correction]}.

\subsection{Gluon Self-energy}
Let us now apply the reduction strategy to the gluon self-energy diagrams. For the tadpole diagram, we simply use Eq.~(\ref{eq:338}) and we find
\beq
\boxed{(d-1)\frac{\Pi_{tad,\perp}(P)}{g^2N}= \frac{(d-1)^3}{d}J_m}\,\,.
\eeq
The ghost loop is also easily treated as one can use Eq.~(\ref{eq:346}) with $m=0$. Upon integration, only one term survives and we find
\beq
\boxed{(d-1)\frac{\Pi_{gh,\perp}(P)}{g^2N}=-\frac{P^2}{4}I_{00}(P)}\,\,.
\eeq
The gluon loop is a bit more cumbersome but doable. We start by expressing Eq.~(\ref{eq:335}) in terms of the auxiliary integrals
\beq
\hat I_{m_1m_2}(P)\equiv\int_Q \sigma\,G_{m_1}(Q)G_{m_2}(R)=\hat I_{m_2m_1}(P)\,.
\eeq
To this purpose, we use Eq.~(\ref{eq:336}), (\ref{eq:337}) and (\ref{eq:338}) to find
\beq
& & (d-1)\frac{\Pi_{gl,\perp}(P)}{g^2N}\nonumber\\
& & \hspace{1.0cm}=\,\left[\left(-\frac{2}{P^2}+\frac{4}{m^2}\right)\hat I_{mm}(P)-\frac{4}{m^2}\hat I_{m0}(P)\right]\left(d-\frac{3}{2}\right)\nonumber\\
& & \hspace{1.5cm}+\,\frac{1}{P^2}\Big[\hat I_{m0}(P)-\hat I_{mm}(P)\Big]\nonumber\\
& & \hspace{1.5cm}-\,\frac{P^2}{2m^4}\Big[\hat I_{mm}(P)+\hat I_{00}(P)-2\hat I_{m0}(P)\Big]\nonumber\\
& & \hspace{1.5cm}-\,\frac{d-1}{d}J_m\,,
\eeq
or
\beq
& & (d-1)\frac{\Pi_{gl,\perp}(P)}{g^2N}\nonumber\\
& & \hspace{1.0cm}=\,\left[\left(-\frac{2}{P^2}+\frac{4}{m^2}\right)\hat I_{mm}(P)-\frac{4}{m^2}\hat I_{m0}(P)\right]\left(d-\frac{3}{2}\right)\nonumber\\
& & \hspace{1.5cm}+\,\frac{P^4+m^4}{P^2m^4}\hat I_{m0}(P)-\frac{P^4+2m^4}{2P^2m^4}\hat I_{mm}(P)-\frac{P^2}{2m^4}\hat I_{00}(P)\nonumber\\
& & \hspace{1.5cm}-\,\frac{d-1}{d}J_m\,.\label{eq:354}
\eeq
Now, the integration of Eq.~(\ref{eq:346}) can be seen as providing a relation between $\hat I_{m0}$, $I_{m0}$ and $J_m$, and also between $\hat I_{00}$ and $I_{00}$:
\beq
\hat I_{m0}(P) & = & -\frac{(P^2+m^2)^2}{4}I_{m0}(P)+\frac{P^2-m^2}{4}J_m\,,\nonumber\\
\hat I_{00}(P) & = & -\frac{(P^2)^2}{4}I_{00}(P)\,.
\eeq
We just need to derive a similar relation for $\hat I_{mm}$. To this purpose we write
\beq
Q^2R^2G(Q)G(R) & = & (1-m^2G(Q))(1-m^2G(R))\nonumber\\
& = & 1-m^2(G(Q)+G(R))+m^4G(Q)G(R)\,,\label{eq:357}
\eeq
and
\beq
(Q\cdot R)G(Q)G(R) & = & \frac{1}{2}\Big[P^2-Q^2-R^2\Big]G(Q)G(R)\nonumber\\
& = & \frac{1}{2}\Big[P^2+2m^2-(Q^2+m^2)-(R^2+m^2)\Big]G(Q)G(R)\nonumber\\
& = & \frac{P^2+2m^2}{2}G(Q)G(R)-\frac{1}{2}G(R)-\frac{1}{2}G(Q)\,,
\eeq
which we iterate once to find
\beq
(Q\cdot R)^2G(Q)G(R) & = & \frac{P^2+2m^2}{2}\left[\frac{P^2+2m^2}{2}G(Q)G(R)-\frac{1}{2}G(R)-\frac{1}{2}G(Q)\right]\nonumber\\
& & -\frac{Q\cdot R}{2}G(R)-\frac{Q\cdot R}{2}G(Q)\,.
\eeq
Combining this with (\ref{eq:357}), we find
\beq
\sigma\,G(Q)G(R) & = & 1+\left[\frac{Q\cdot R}{2}+\frac{P^2-2m^2}{4}\right]G(R)+\left[\frac{Q\cdot R}{2}+\frac{P^2-2m^2}{4}\right]G(Q)\nonumber\\
& & +\left(m^4-\frac{(P^2+2m^2)^2}{4}\right)G(Q)G(R)\,.
\eeq
Upon using $P+Q+R=0$, this becomes
\beq
\sigma\,G(Q)G(R) & = & \left[-\frac{P\cdot R}{2}+\frac{P^2}{4}\right]G(R)+\left[-\frac{P\cdot Q}{2}+\frac{P^2}{4}\right]G(Q)\nonumber\\
& & -\frac{P^2(P^2+4m^2)}{4}G(Q)G(R)\,.
\eeq
Upon integration this gives
\beq
\hat I_{mm}(P)=-\frac{P^2(P^2+4m^2)}{4}I_{mm}(P)+\frac{P^2}{2}J_m\,.
\eeq
Using this and the other reduction formulae in Eq.~(\ref{eq:354}), we find
\beq
& & \Bigg|\Bigg|(d-1)\frac{\Pi_{gl,\perp}(P)}{g^2N}\nonumber\\
& & \hspace{1.0cm}=\,\left[d\frac{P^2}{m^2}-\frac{7P^4+5P^2m^2+m^4}{4P^2m^2}+\frac{1}{d}\right]J_m\nonumber\\
& & \hspace{1.0cm}+\,(P^2+4m^2)\left[\left(\frac{1}{2}-\frac{P^2}{m^2}\right)d+\frac{P^4+12P^2m^2-4m^4}{8m^4}\right]I_{mm}\nonumber\\
& & \hspace{1.0cm}+\,\frac{(P^2+m^2)^2}{m^2}\left[d-\frac{P^4+6P^2m^2+m^4}{4P^2m^2}\right]I_{m0}\nonumber\\
& & \hspace{1.0cm}+\,\frac{(P^2)^3}{8m^4}I_{00}\,.
\eeq
When adding the contributions from the other two diagrams, we arrive at
\beq
& & \Bigg|\Bigg|(d-1)\frac{\Pi_{\perp}(P)}{g^2N}\nonumber\\
& & \hspace{1.0cm}=\,\left[d^2+\left(\frac{P^2}{m^2}-3\right)d-\frac{7P^4-7P^2m^2+m^4}{4P^2m^2}\right]J_m\nonumber\\
& & \hspace{1.0cm}+\,(P^2+4m^2)\left[\left(\frac{1}{2}-\frac{P^2}{m^2}\right)d+\frac{P^4+12P^2m^2-4m^4}{8m^4}\right]I_{mm}\nonumber\\
& & \hspace{1.0cm}+\,\frac{(P^2+m^2)^2}{m^2}\left[d-\frac{P^4+6P^2m^2+m^4}{4P^2m^2}\right]I_{m0}\nonumber\\
& & \hspace{1.0cm}+\,\left(\frac{(P^2)^2}{2m^4}-1\right)\frac{P^2}{4}I_{00}\,.\label{eq:264}
\eeq

\section{Final Result}
The integrals $J_m$, $I_{mm}$, $I_{m0}$ and $I_{00}$ have known expressions, see Appendix B, from which one could complete the calculation and obtain explicit one-loop expressions for the self-energies. Here, we focus on the divergent parts. The final expression for the self-energies is given in Appendix B.

\subsection{Extracting the Divergences}
 Using
\beq
J_m=-\frac{m^2}{16\pi^2}\frac{1}{\epsilon}+{\cal O}(\epsilon^0) \quad {\rm and} \quad I_{m_1m_2}(P)=\frac{1}{16\pi^2}\frac{1}{\epsilon}+{\cal O}(\epsilon^0)\,,
\eeq
we find that the divergence of the ghost self-energy is
\beq
\Sigma_{\rm div} & \!\!\!=\!\!\! & \frac{g^2N}{16\pi^2}\frac{1}{\epsilon}\left[\frac{(P^2)^2}{4m^2}-\frac{(P^2+m^2)^2}{4m^2}+\frac{P^2-m^2}{4m^2}(-m^2)\right]\nonumber\\
& \!\!\!=\!\!\! & -\frac{3}{4}\frac{g^2N}{16\pi^2}\frac{P^2}{\epsilon}\,.
\eeq
We note that the divergence vanishes as $P^2$, as expected from the general discussion of the previous chapter. The same can be checked for the finite contribution to the self-energy. The self-energy enters the renormalized dressing function as
\beq
{\cal F}(P)=\frac{P^2}{Z_cP^2+\Sigma(P)}\,,
\eeq
where $Z_cP^2$ is the tree-level part after rescaling of the ghost fields. It follows
\beq
\boxed{Z_c=1+\frac{3}{4}\frac{g^2N}{16\pi^2}\frac{1}{\epsilon}+{\cal O}(\epsilon^0)}\,\,.
\eeq
As for the gluon self-energy, we find the divergence
\beq
& & 3\frac{\Pi^{\perp}_{\rm div}(P)}{g^2N}\nonumber\\
& & \hspace{1.0cm}=\,\left[16+\left(\frac{P^2}{m^2}-3\right)4-\frac{7P^4-7P^2m^2+m^4}{4P^2m^2}\right]\frac{-m^2}{16\pi^2\epsilon}\nonumber\\
& & \hspace{1.0cm}+\,(P^2+4m^2)\left[\left(\frac{1}{2}-\frac{P^2}{m^2}\right)4+\frac{P^4+12P^2m^2-4m^4}{8m^4}\right]\frac{1}{16\pi^2\epsilon}\nonumber\\
& & \hspace{1.0cm}+\,\frac{(P^2+m^2)^2}{m^2}\left[4-\frac{P^4+6P^2m^2+m^4}{4P^2m^2}\right]\frac{1}{16\pi^2\epsilon}\nonumber\\
& & \hspace{1.0cm}+\,\left(\frac{(P^2)^2}{2m^4}-1\right)\frac{P^2}{4}\frac{1}{16\pi^2\epsilon}\nonumber\\
& & \hspace{1.0cm}=\frac{9m^2-26P^2}{4}\frac{1}{16\pi^2}\frac{1}{\epsilon}
\eeq
and thus
\beq
\Pi^{\perp}_{\rm div}(P)=\left(\frac{3}{4}m^2-\frac{13}{6}P^2\right)\frac{g^2N}{16\pi^2}\frac{1}{\epsilon}\,.
\eeq
The self-energy enters the renormalized gluon propagator as
\beq
{\cal G}(P)=\frac{1}{Z_AP^2+Z_AZ_{m^2}m^2+\Pi^\perp(P)}\,.
\eeq
It follows that
\beq
\boxed{Z_A=1+\frac{13}{6}\frac{g^2N}{16\pi^2}\frac{1}{\epsilon}+{\cal O}(\epsilon^0)}\,\,,
\eeq
and
\beq
\boxed{Z_AZ_{m^2}=1-\frac{3}{4}\frac{g^2N}{16\pi^2}\frac{1}{\epsilon}+{\cal O}(\epsilon^0)}\,\,,
\eeq
from which we deduce that
\beq
\boxed{Z_{m^2}=1-\frac{35}{12}\frac{g^2N}{16\pi^2}\frac{1}{\epsilon}+{\cal O}(\epsilon^0)}\,\,.
\eeq
Notice that
\beq
Z_cZ_AZ_{m^2}={\cal O}(\epsilon^0)\,,
\eeq
in agreement with the non-renormalization theorem derived in the previous chapter. From the other non-renormalization theorem stating that $Z_{g^2}Z_AZ_c^2$ is finite, we deduce that
\beq
\boxed{Z_{g^2}=1-\frac{11}{3}\frac{g^2N}{16\pi^2}\frac{1}{\epsilon}+{\cal O}(\epsilon^0)}\,\,,
\eeq
which is the expected result for YM theory.

\subsection{Comparison to the Lattice}
To fix the finite parts of the renormalization factors, one needs to specify a renormalization scheme. One possibility is to impose a certain number (here $4$) of renormalization conditions which can be seen as a choice of definition of the renormalized parameters (mass and coupling) and normalization of the fields.

Once the renormalization factors are fully determined they can be used in the renormalized propagators to obtain explicitly finite expressions. These expressions are then fitted to the lattice data by scanning the possible values of $m$ and $g$, while allowing for adjustable normalizations of the ghost and gluon propagators since the propagators computed in different schemes (the lattice scheme versus whatever continuum scheme we decide to use) differ by an overall factor.

\begin{figure}[h]
\begin{center}
\includegraphics[height=0.4\textheight]{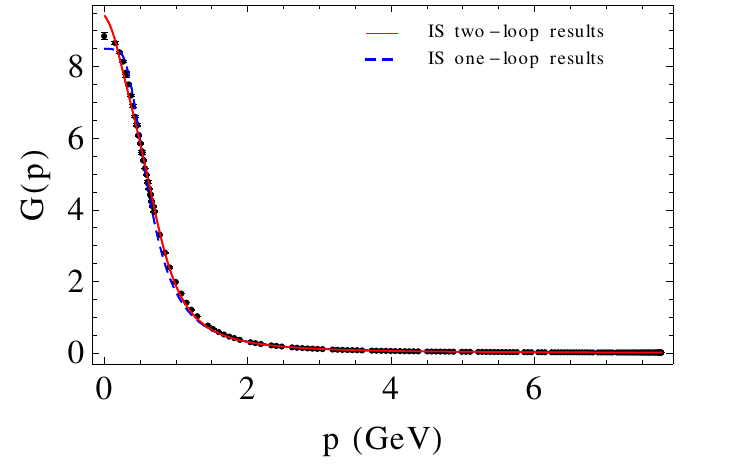}\\
\includegraphics[height=0.4\textheight]{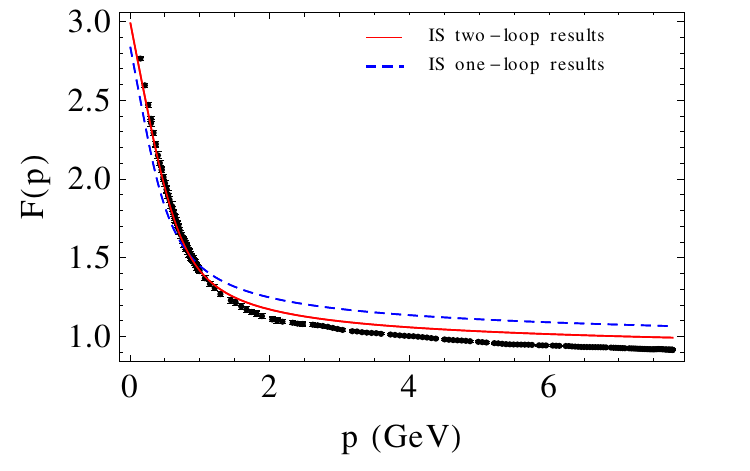}
\caption{Fits of the SU($3$) Landau-gauge lattice gluon propagator and ghost dressing function using one- and two-loop expression as computed within the CF model \cite{Gracey:2019xom}. The original one-loop fits can be found in Refs.~\cite{Tissier:2010ts,Tissier:2011ey}.}\label{fig:props}
\end{center}
\end{figure}

The results of the fitting procedure in the SU($3$) case are shown in Fig.~\ref{fig:props},\footnote{Strictly speaking, these comparisons do not use strict perturbation theory but rather renormalization group improved perturbation theory. This is needed in particular to have a good control on the large momentum tails, as we discuss in the next lecture.} together with similar fits using improved two-loop expressions (the derivation of which goes far beyond the scope of these lectures). The agreement is already quite good at one-loop order and improves when including higher order corrections. These results extend to other correlation functions. This points to the good adequacy of the Curci-Ferrari model in coping with the deficiencies of the Faddeev-Popov gauge fixing and opens the way to a study of the infrared properties of YM/QCD theories.

\chapter{Renormalization Group Flow}

In this final lecture, we deepen our understanding of the Curci-Ferrari model by analysing its renormalization group (RG) flow trajectories, which tell us in particular what to expect from the model at large or low energies.

\section{Renormalization Group}
The need for eliminating the divergences introduces an arbitrary scale $\mu$ in the problem, known as the renormalization scale. Usually, this scale enters the renormalization conditions, a set of conditions on the correlation/vertex functions that define the renormalized parameters and the normalizations of the fields. We shall give some examples of renormalization conditions below. In the special case of the minimal subtraction scheme in dimensional regularization, the renormalization scale enters through the relation
\beq
g^2_B=\mu^{2\epsilon} Z_{g^2} g^2\label{eq:gB}
\eeq
between the bare coupling $g^2_B$ which is of mass dimension $4-d=2\epsilon$ and the dimensionless renormalized coupling $g$. Notice that the relation between the bare mass and the renormalized mass does not require the presence of $\mu$:
\beq
m^2_B=Z_{m^2}m^2\label{eq:mB}
\eeq
since the bare mass if already of mass dimension $2$.

\subsection{Renormalized Vertex Functions}
In any case, the renormalized vertex functions become functions of $\mu$:
\beq
\Gamma^{(n)}({\{p_i\},m^2(\mu),g^2(\mu);\mu})=Z^{n/2}(\mu,\epsilon)\Gamma_B(\{p_i\},m^2_B,g^2_B;\epsilon)
\eeq
and only ratios of vertex functions at two different configuration of momenta are comparable across different renormalization schemes. 

One inconvenience of the appearance of this extra scale $\mu$ is that whenever there is a too large separation between that scale and another scale of the problem, they may appear large logarithms that spoil the validity of the perturbative expansion, even when the coupling is small. In YM theories, this is typically what happens in the ultraviolet tails of the correlation functions, that is for $p^2_i\sim p^2$ and $p^2/\mu^2\gg 1$. 

The way to handle this problem is to take the ratio of the above equation at two different values of $\mu$, a running value which we still denote as $\mu$ and a fixed value denoted $\mu_0$:
\beq
\Gamma^{(n)}({\{p_i\},m^2_0,g^2_0;\mu_0})=\frac{Z^{n/2}(\mu_0,\epsilon)}{Z^{n/2}(\mu,\epsilon)}\Gamma^{(n)}({\{p_i\},m^2(\mu),g^2(\mu);\mu})\,.
\eeq
Then, one can choose the arbitrary scale $\mu$ in the right-hand side to be the large scale $p$:
\beq
\Gamma^{(n)}({\{p_i\},m^2_0,g^2_0;\mu_0})=\frac{Z^{n/2}(\mu_0,\epsilon)}{Z^{n/2}(p,\epsilon)}\Gamma^{(n)}({\{p_i\},m^2(p),g^2(p);p})\,.\label{eq:ddd}
\eeq
The left-hand side can still not be determined perturbatively for $p\gg\mu_0$ due to the large logarithms. But the  right-hand side can since such large logarithms are absent. 

Of course, this procedure works provided we are able to determine the running parameters $m(\mu)$ and $g(\mu)$, as well as the (finite) prefactor
\beq
z(\mu_0,\mu)\equiv\frac{Z(\mu_0,\epsilon)}{Z(\mu,\epsilon)}\,.\label{eq:z}
\eeq

\subsection{$\beta$-functions and Anomalous Dimensions}
The running of the parameters is most easily obtained from the integration of the $\beta$ functions defined as
\beq
\beta_{g^2}\equiv\mu\frac{\partial g^2}{\partial\mu} \quad {\rm and} \quad \beta_{m^2}\equiv\mu\frac{\partial m^2}{\partial\mu}\,.
\eeq
The latter can be evaluated by noting that the bare parameters (\ref{eq:gB}) and (\ref{eq:mB}) should not depend on the renormalization scale. Taking the logarithm of these definitions and then a $\mu$-derivative, one finds the relations
\beq
0=\mu\frac{\partial \ln m^2}{\partial\mu}+\mu\frac{\partial\ln Z_{m^2}}{\partial\mu} \quad {\rm and} \quad 0= 2\epsilon+\mu\frac{\partial \ln g^2}{\partial\mu}+\mu\frac{\partial\ln Z_{g^2}}{\partial\mu}\,,
\eeq
which allow us to obtain the $\beta$-functions as
\beq
\boxed{\beta_{m^2}=-m^2\gamma_{m^2}} \quad {\rm and} \quad \boxed{\beta_{g^2}=-(2\epsilon+\gamma_{g^2})g^2}\,\,,\label{eq:bg}
\eeq
in terms of the anomalous dimensions
\beq
\gamma_{m^2}\equiv\mu\frac{\partial\ln Z_{m^2}}{\partial\mu} \quad {\rm and} \quad \gamma_{g^2}\equiv\mu\frac{\partial\ln Z_{g^2}}{\partial\mu}\,,
\eeq
that relate directly to the renormalization factors.

Similarly, by taking the logarithm of (\ref{eq:z}) and then a $\mu$-derivative, one finds
\beq
\mu\frac{\partial \ln z(\mu_0,\mu)}{\partial\mu}=-\mu\frac{\partial\ln Z}{\partial\mu}\equiv-\gamma\,,
\eeq
with $\gamma$ tha anomalous dimension associated to the field renormalization (one for each type of field). Integrating this back, one obtains
\beq
\boxed{z(\mu_0,\mu)=\exp\left\{\int_\mu^{\mu_0}\frac{d\nu}{\nu}\,\gamma(\nu)\right\}=\exp\left\{\int_{g^2(\mu)}^{g^2_0}\frac{dg}{\beta_{g^2}}\,\gamma(g^2)\right\}}\,\,.\label{eq:zz}
\eeq

\section{Ultraviolet Flow}
Let us first consider the RG flow in the ultraviolet (UV). At one-loop order, it does not depend on the considered scheme, so we may use the minimal subtraction scheme for which
\beq
Z_{g^2} & \!\!\!=\!\!\! & 1-\frac{11}{3}\frac{g^2N}{16\pi^2}\frac{1}{\epsilon}\,,\\
Z_{m^2} & \!\!\!=\!\!\! & 1-\frac{35}{12}\frac{g^2N}{16\pi^2}\frac{1}{\epsilon}\,,\\
Z_A & \!\!\!=\!\!\! & 1+\frac{13}{6}\frac{g^2N}{16\pi^2}\frac{1}{\epsilon}\,,\\
Z_c & \!\!\!=\!\!\! & 1+\frac{3}{4}\frac{g^2N}{16\pi^2}\frac{1}{\epsilon}\,.
\eeq

\subsection{Running Coupling}
We then have
\beq
\gamma_{g^2} & \!\!\!=\!\!\! & \mu\frac{\partial\ln Z_{g^2}}{\partial\mu}=\frac{\mu}{Z_{g^2}}\frac{\partial Z_{g^2}}{\partial\mu}\nonumber\\
& \!\!\!=\!\!\! & \mu\frac{\partial Z_{g^2}}{\partial\mu}+\mbox{higher orders}\nonumber\\
& \!\!\!=\!\!\! & -\frac{11}{3}\frac{\beta_{g^2}N}{16\pi^2}\frac{1}{\epsilon}+{\cal O}(g^4)\,.\label{eq:gg}
\eeq
It may seem that we have reached a circular reasoning since we need $\beta_{g^2}$ to determine $\gamma_{g^2}$ but it is precisely from $\gamma_{g^2}$ that we planned to evaluate $\beta_{g^2}$ in the first place. There is no problem, however, because, to the order we are computing, we only need the leading contribution to $\beta_{g^2}$ which does not depend on $\gamma_{g^2}$ and is given explicitly in Eq.~(\ref{eq:bg}) . Moreover, this contribution comes with a welcomed factor of $\epsilon$ that cancels the annoying pole in Eq.~(\ref{eq:gg}). All in all, we obtain the UV finite, one-loop finite anomalous dimension
\beq
\gamma_{g^2}=\frac{11}{3}\frac{g^2N}{16\pi^2}\,,
\eeq
and thus the one-loop finite beta function
\beq
\boxed{\beta_{g^2}=-\frac{22}{3}\frac{g^4N}{16\pi^2}}\,\,.
\eeq
This is the well known one-loop YM beta function. It rewrites as
\beq
-\frac{dg^2}{g^4}=\frac{11N}{48\pi^2}d\ln\mu^2\,,
\eeq
which integrates to
\beq
\frac{1}{g^2(\mu)}-\frac{1}{g_0^2}=\frac{11N}{48\pi^2}\ln\frac{\mu^2}{\mu^2_0}\,.
\eeq
Introducing the scale $\mu=\Lambda$ such that $1/g^2(\Lambda)=0$, we have
\beq
-\frac{1}{g_0^2}=\frac{48\pi^2}{11}\ln\frac{\Lambda^2}{\mu_0^2}\,.
\eeq
which we can subtract from the expression above to arrive at
\beq
\frac{1}{g^2(\mu)}=\frac{11N}{48\pi^2}\ln\frac{\mu^2}{\Lambda^2}
\eeq
or
\beq
\boxed{g^2(\mu)=\frac{48\pi^2}{11N}\frac{1}{\ln\frac{\mu^2}{\Lambda^2}}}\,\,.
\eeq
The running goes to $0$ as $\mu\gg\Lambda$ (asymptotic freedom) and diverges as $\mu\to\Lambda^+$ (IR Landau pole).
 
\subsection{Running Mass}
As for the running mass, a similar calculation leads to
\beq
\gamma_{m^2}=\frac{35}{6}\frac{g^2N}{16\pi^2}\,,
\eeq
and thus
\beq
\beta_{m^2}=-\frac{35}{6}\frac{g^2N}{16\pi^2}m^2\,.
\eeq
From the ratio of the $\beta$-functions, it follows that
\beq
\frac{\beta_{m^2}}{\beta_{g^2}}=\frac{dm^2}{dg^2}=\frac{35}{44}\frac{m^2}{g^2}\,,
\eeq
or
\beq
d\ln m^2=\frac{35}{44}d\ln g^2\,.
\eeq
This integrates to
\beq
\boxed{\frac{m^2(\mu)}{m^2_0}=\left(\frac{g^2(\mu)}{g^2_0}\right)^{35/44}}\,\,.
\eeq
In particular, the mass goes to $0$ in the UV! In a sense, this shows that the family of CF models and the FP are not distinguishable microscopically. One needs an extra information to know which one is the relevant one has one includes fluctuations.

\subsection{Field Anomalous Dimensions}
We can also compute the field anomalous dimensions
\beq
\gamma_c=-\frac{3}{2}\frac{g^2N}{16\pi^2} \quad {\rm and} \quad \gamma_A=-\frac{13}{3}\frac{g^2N}{16\pi^2}\,,
\eeq
and thus from (\ref{eq:zz})
\beq
z_c(\mu_0,\mu) & \!\!\!=\!\!\! & \exp\left\{\int_{g^2(\mu)}^{g^2_0}\frac{dg^2}{\beta_{g^2}}\,\gamma_c(g^2)\right\}\nonumber\\
& \!\!\!=\!\!\! & \exp\left\{\frac{9}{44}\int_{g^2(\mu)}^{g^2_0}\frac{dg^2}{g^2}\right\}=\left(\frac{g^2(\mu)}{g^2_0}\right)^{-9/44}\,,
\eeq
and, similarly,
\beq
z_A(\mu_0,\mu)=\left(\frac{g^2(\mu)}{g^2_0}\right)^{-13/22}\,.
\eeq

\section{Infrared Flow}
The behavior of the renormalization group trajectories in the infrared depends on the scheme. In principle, changing from one scheme to another corresponds to mapping the trajectories between the two schemes. In practice however, a given scheme may not be able to explore all possible trajectories of the model. Interestingly enough, within the CF model, while the minimal subtraction scheme leads necessarily to trajectories with a Landau pole, one can construct schemes which contain trajectories without Landau pole in addition to those that present a Landau pole. The Infrared Safe Scheme introduced by Tissier and Wschebor is one example of such schemes.

\subsection{Infrared Safe Scheme}
The renormalization factors are fixed via the conditions
\beq
{\cal D}^{-1}(p=\mu;\mu)=\mu^2 \quad {\rm and} \quad {\cal G}^{-1}(p=\mu;\mu)=\mu^2+m^2\,,\label{eq:cond1}
\eeq
supplemented by
\beq
Z_{g^2} Z_A Z_c^2=1 \quad {\rm and} \quad Z_{m^2}Z_AZ_c=1\,.\label{eq:cond2}
\eeq
We note that the last two conditions are only possible because of the existence of two non-renormalization theorems that constraint the left-hand sides of these two conditions to be finite. The present just amounts to choosing this finite quantities to be equal to $1$.

The first condition in Eq.~(\ref{eq:cond1}) fixes $Z_c$ while the second determines $Z_A$ in terms of $Z_AZ_{m^2}$, but since the second condition in Eq.~(\ref{eq:cond2}) determines $Z_{m^2}Z_A$ in terms of the already determined $Z_c$, this also determines $Z_A$ and $Z_{m^2}$. Finally, from the first condition in Eq.~(\ref{eq:cond2}), one gets $Z_{g^2}$.

\subsection{$\beta$-functions}
A peculiarity of the IR-safe scheme is that the $\beta$-functions are entirely determined from the field anomalous dimensions. Indeed, taking the logarithm of the third and fourth renormalization conditions  followed by a $\mu$-derivative, one finds \cite{Tissier:2010ts}
\beq
\gamma_{g^2}+\gamma_A+2\gamma_c=0 \quad {\rm and} \quad \gamma_{m^2}+\gamma_A+\gamma_c=0\,,
\eeq
from which one easily deduces that
\beq
\boxed{\beta_{g^2}=g^2\left(\gamma_A+2\gamma_c\right)} \quad {\rm and} \quad \boxed{\beta_{m^2}=m^2(\gamma_A+\gamma_c)}\,\,.\label{eq:rel}
\eeq
The anomalous dimensions are easily computed from the renormalization factors $Z_A$ and $Z_c$. In the IR-safe scheme, one finds
\eq{\label{eq:gammac}
\gamma_c=-\frac{\lambda}{2t^2} \Big[2t^2+2t -t^3 \ln t+(t+1)^2 (t-2) \ln (t+1)\Big]\,,
}
where we denote $t\equiv\mu^2/m^2$, and
\eq{
\label{eq:gammaA}
\gamma_A&=\frac{\lambda}{6t^3}  \Bigg[-17 t^3+74t^2-12t+t^5 \ln t\nn
&-(t-2)^2 (2 t-3)
  (t+1)^2 \ln (t+1) \nn
&-t^{3\over2} \sqrt{t+4} \left(t^3-9
   t^2+20 t-36\right)\ln\!\left(\frac{\sqrt{t+4}-\sqrt{t}}{\sqrt{t+4}+\sqrt{t}}\right)\!\!\Bigg]\,.
}

\subsection{RG Trajectories}
When studying the flow trajectories, it is more convenient to work with dimensionless parameters $g$ and $\smash{\tilde m^2\equiv m^2/\mu^2=1/t}$ such that
\beq
\frac{\beta_{\tilde m^2}}{\tilde m^2}=\frac{\beta_{m^2}}{m^2}-2\,,
\eeq
and thus
\beq
\beta_{\tilde m^2}=\tilde m^2(\gamma_A+\gamma_c-2)\,.
\eeq
In particular, we can look for the existence of fixed points of the RG flow such that $\beta_{g^2}=0$ and $\beta_{\tilde m^2}=0$. In addition to the UV fixed point corresponding to $\smash{g^2=0}$, $\smash{\tilde m^2=0}$, one finds a non-trivial fixed point
\beq
\gamma_A+2\gamma_c=0 \quad {\rm and} \quad \gamma_A+\gamma_c=2\,.
\eeq
The first equation determines a particular value of $\smash{\tilde m^2=\tilde m^2_\star}$ while the second equation allow one to determine the corresponding $g^2_\star$. One finds
\beq
\tilde m^2_\star\approx 14.18 \quad {\rm and} \quad \frac{g^2_\star N}{16\pi^2}\approx 16.11\,.
\eeq
The value of $g^2_\star N(16\pi^2)$ is too large for this result to be taken for granted within a one-loop calculation. It should be mentioned however that a two-loop calculation leads instead to  $g^2_\star N(16\pi^2)\approx 1.64$ a still non-perturbative but more moderate value of the coupling.

\begin{figure}[t]
\begin{center}
\includegraphics[height=0.4\textheight]{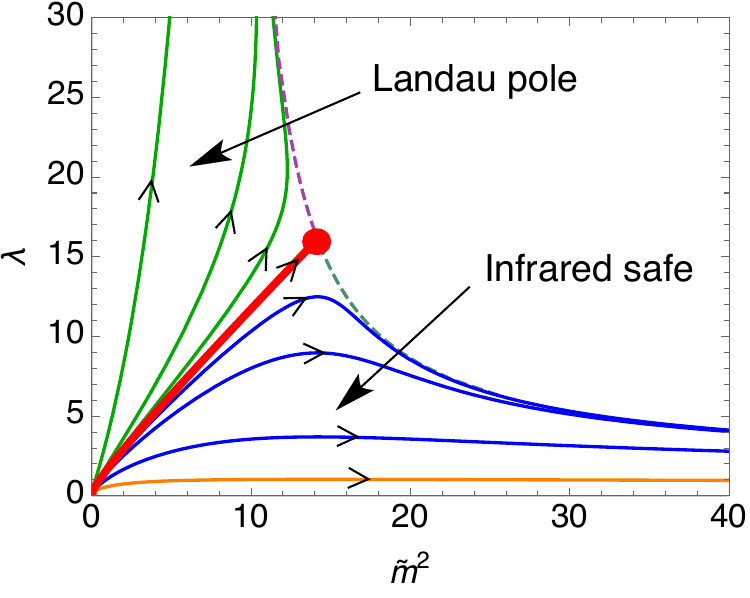}\\
\caption{Renormalization Group flow of the Curci-Ferrari model. Here $\smash{\lambda\equiv g^2N/(16\pi^2)}$. Plot taken from Ref.~\cite{Reinosa:2017qtf} but the original RG discussion can be found in Ref.~\cite{Tissier:2010ts}.}\label{fig:flow}
\end{center}
\end{figure}

In any case, it is interesting to note that the UV and IR fixed points are connected by one particular flow trajectory that plays the role of a separatrix, see Fig.~\ref{fig:flow}. To the left of this trajectory, all trajectories display a Landau pole as obtained also within the minimal subtraction scheme and as it happens also within the Faddeev-Popov case, which is a actually contained in this region since it corresponds to $\smash{\tilde m^2=0}$. On the contrary, to the right of this separatrix, the trajectories are defined down to $\mu=0$ and are thus free of a Landau pole. The IR-fixed point is not stable w.r.t to perturbations away from the separatrix. In this case the actual IR fixed point corresponds to $\tilde m^2=\infty$ and $g^2=0$. This can be seen by expanding the anomalous dimension for large $\tilde m^2$. One finds
\beq
\gamma_A\approx \frac{g^2N}{48\pi^2} \quad {\rm while} \quad \gamma_c\approx 0\,.
\eeq
It follows that
\beq
\frac{\beta_{g^2}}{g^2}\approx\frac{\beta_{m^2}}{m^2}\approx  \frac{g^2N}{48\pi^2}\,,
\eeq
from which we deduce that
\beq
\boxed{g^2\sim-\frac{48\pi^2}{N}\frac{1}{\ln\frac{\mu}{\mu_0}}\to 0}\,,
\eeq
and
\beq
\boxed{m^2\propto g^2\to 0}\,\,!!!
\eeq
with $m^2/\mu^2\to \infty$.

\subsection{Propagators}
To compute the propagators, we use Eq.~(\ref{eq:ddd}), with Eq.~(\ref{eq:zz}) and $n=2$:
\beq
{\cal D}^{-1}(p;m_0^2,g_0^2,\mu_0) & \!\!\!=\!\!\! &\exp\left\{\int_\mu^{\mu_0}\frac{d\nu}{\nu}\,\gamma_c(\nu)\right\}{\cal D}^{-1}(p;m(p)^2,g(p)^2,\mu=p)\,,\nonumber\\\\
{\cal G}^{-1}(p;m_0^2,g_0^2,\mu_0) & \!\!\!=\!\!\! &\exp\left\{\int_\mu^{\mu_0}\frac{d\nu}{\nu}\,\gamma_A(\nu)\right\}{\cal G}^{-1}(p;m(p)^2,g(p)^2,\mu=p)\,.\nonumber\\
\eeq
But now, from the very choice of renormalization conditions, the propagators at $\mu=p$ have very simple expressions
\beq
{\cal D}^{-1}(p;m(p)^2,g(p)^2,\mu=p)=p^2\,,
\eeq
and
\beq 
{\cal G}^{-1}(p;m(p)^2,g(p)^2,\mu=p)=p^2+m^2(p)\,.
\eeq
Moreover, the anomalous dimension can be entirely expressed in terms of the $\beta$-functions by inverting the relations (\ref{eq:rel}):
\beq
\gamma_c=\frac{\beta_{g^2}}{g^2}-\frac{\beta_{m^2}}{m^2} \quad {\rm and} \quad \gamma_A=2\frac{\beta_{m^2}}{m^2}-\frac{\beta_{g^2}}{g^2}\,,
\eeq
so
\beq
\exp\left\{\int_\mu^{\mu_0}\frac{d\nu}{\nu}\,\gamma_c(\nu)\right\} & = & \exp\left\{\int_\mu^{\mu_0}\frac{d\nu}{\nu}\,\left(\frac{\beta_{g^2}}{g^2}-\frac{\beta_{m^2}}{m^2}\right)\right\}\nonumber\\
& = & \exp\left\{\int_\mu^{\mu_0}d\nu\,\left(\frac{d\ln g^2}{d\nu}-\frac{d\ln m^2}{d\nu}\right)\right\}\nonumber\\
& = & \exp\left\{\ln\frac{g^2_0}{g^2(\mu)}-\ln\frac{m^2_0}{m^2(\mu)}\right\}=\frac{g_0^2}{g^2(\mu)}\frac{m^2(\mu)}{m^2_0}\,,\nonumber\\
\eeq
and, similarly,
\beq
\exp\left\{\int_\mu^{\mu_0}\frac{d\nu}{\nu}\,\gamma_A(\nu)\right\}=\frac{g^2(\mu)}{g^2_0}\frac{m^4_0}{m^4(\mu)}\,.
\eeq
Putting all the pieces together, we find that
\beq
{\cal D}(p;m_0^2,g_0^2,\mu_0) & \!\!\!=\!\!\! & \frac{m^2_0}{g^2_0}\frac{g^2(p)}{m^2(p)}\frac{1}{p^2}\,,\\
{\cal G}(p;m_0^2,g_0^2,\mu_0) & \!\!\!=\!\!\! & \frac{g^2_0}{m^2_4}\frac{m^4(p)}{g^2(p)}\frac{1}{p^2+m^2(p)}\,.
\eeq
Notice that, in the UV, using the running derived above for $m^2$, we find
\beq
{\cal D}(p;m_0^2,g_0^2,\mu_0) & \!\!\!\propto\!\!\! & \frac{(g^2(p))^{9/44}}{p^2}\,,\\
{\cal G}(p;m_0^2,g_0^2,\mu_0) & \!\!\!\propto\!\!\! & \frac{(g^2(p))^{13/22}}{p^2}\,,
\eeq
in agreement with the anomalous dimensions that we determined above. As for the IR, since $m^2\propto g^2$, we find that 
\beq
{\cal D}(p;m_0^2,g_0^2,\mu_0) & \!\!\!\propto\!\!\! & \frac{1}{p^2}\,,\\
{\cal G}(p;m_0^2,g_0^2,\mu_0) & \!\!\!\to\!\!\! & {\rm constant}\,.
\eeq
So, even though the mass parameter runs to $0$ both in the UV and in the IR, the corresponding gluon propagator displays a massive behavior!

\section{Comparison to the Lattice}
\begin{figure}[t]
\begin{center}
\includegraphics[height=0.35\textheight]{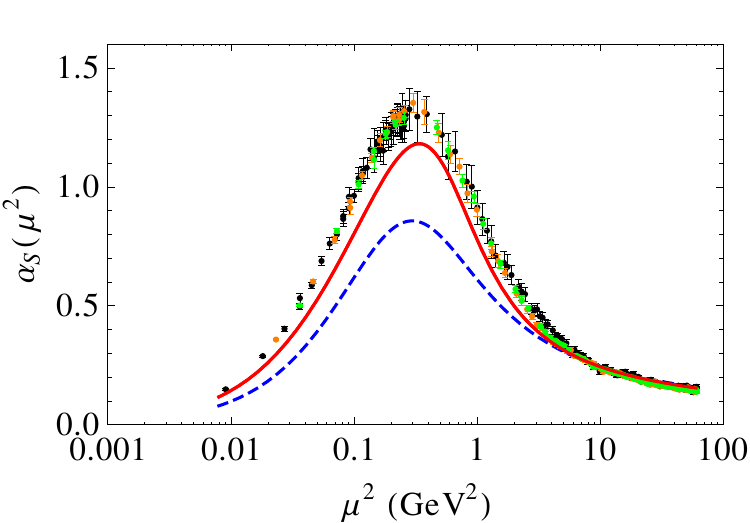}\\
\caption{Running coupling as obtained from the CF model in the infrared-safe scheme at one-loop (dashed) and two-loop (plain) order \cite{Gracey:2019xom}, mapped back to the lattice scheme for comparison. We note that the perturbative expansion parameter is $g^2N/(16\pi^2)=\alpha_SN/(4\pi)$ and does not exceed $1$.}\label{fig:alphaS}
\end{center}
\end{figure}

It is interesting to look more closely at the running coupling in the IR-safe scheme. Since $Z_{g^2}(\mu)Z_A(\mu) Z_c^2(\mu)=1$, we can write
\beq
g^2(\mu)=\mu^{-2\epsilon}g_B^2 Z_A(\mu) Z_c^2(\mu)\,,
\eeq
where $g_B^2$ is the bare coupling. Since the latter does not depend on $\mu$, this implies that
\beq
\frac{g^2(\mu)}{g^2(\mu)}=\left(\frac{\mu}{\nu}\right)^{-2\epsilon}\frac{Z_A(\mu)}{Z_A(\nu)}\frac{Z^2_c(\mu)}{Z^2_c(\nu)}\,.
\eeq
But the ratios of renormalization factors can be written as ratios of renormalized propagators at any value of $p$:
\beq
\frac{g^2(\mu)}{g^2(\nu)}=\frac{{\cal G}(p,m^2(\nu),g^2(\nu),\nu)}{{\cal G}(p,m^2(\mu),g^2(\mu),\mu)}\left(\frac{{\cal D}(p,m^2(\nu),g^2(\nu),\nu)}{{\cal D}(p,m^2(\mu),g^2(\mu),\mu)}\right)^2\,,
\eeq
where we could take the limit $\epsilon\to 0$ since only renormalized quantities appear. Choosing now $\smash{\mu=p}$ and $\smash{\nu=\mu_0}$, we find
\beq
\frac{g^2(p)}{g^2_0}=\frac{{\cal G}(p,m^2_0,g^2_0,\mu_0)}{{\cal G}(p,m^2(p),g^2(p),p)}\left(\frac{{\cal D}(p,m^2_0,g^2_0,\mu_0)}{{\cal D}(p,m^2(p),g^2(p),p)}\right)^2\,,
\eeq
or using the renrmalization conditions
\beq
g^2(p)=g^2_0\,(p^2+m^2(p))\,{\cal G}(p,m^2_0,g^2_0,\mu_0){\cal F}^2(p,m^2_0,g^2_0,\mu_0)\,.
\eeq
In the lattice renormalization, one has instead
\beq
g^2_{\rm latt}(p)= g^2_{0,\rm latt}\,p^2\,{\cal G}(p,m^2_0,g^2_0,\mu_0){\cal F}^2(p,m^2_0,g^2_0,\mu_0)\,.
\eeq
The difference can be explained form the fact that the renormalization scheme differs in that ${\cal G}^{-1}(p=\mu;\mu)=\mu^2$ in this case and thus the gluon renormalization factors differ by $Z_A/Z_A^{\rm latt}=(\mu^2+m^2(\mu))/\mu^2$. The running coupling as obtained from the CF model is compared to the lattice one in Fig.~\ref{fig:alphaS}.

\chapter{Concluding Remarks}
In these lectures, we have reviewed the status of the Curci-Ferrari model as a phenomenological approach for Landau gauge-fixed YM theories in the IR Euclidean domain. 

More precisely, after discussing some of the general properties of the model, we have explained in full detail the evaluation of the one-loop two-point correlators and we have shown their comparison to the corresponding lattice simulations. We have also discussed the flow structure of the model, in particular the presence of Landau-pole-free RG trajectories along which the running coupling never becomes too large. These are precisely the trajectories that allow one to reproduce the lattice data. It is also worth mentioning that the CF model emerges from the same UV fixed point as the FP model. Although identical at a microscopical level, the two types of theories differ at large distances.

Beyond the evaluation of correlation functions, it is also possible to evaluate observables within the CF model. One example is the Polyakov loop which is relevant at finite temperature as a gauge-invariant order parameter for the YM confinement/deconfinement transition. The results obtained within the CF model are quite remarkable and confirm that some of the IR properties of Euclidean YM theory can be described perturbatively \cite{vanEgmond:2021jyx,Surkau:2024zfb}.

In the case of QCD, perturbative methods cannot be used in the IR but the CF model remains a valuable tool since it possesses two small expansion parameters that allow one to keep a control on the error in the evaluation of correlators/observables \cite{Pelaez:2020ups,Pelaez:2022rwx}.

\appendix

\chapter{Proof of Renormalizability}
Let us here discuss the renormalizability of the Curci-Ferrari model. The first step is the determination of the superficial degree of divergence associated to a given diagram.

\section{Superficial Degree of Divergence}
Consider a 1PI diagram and denote $L$ the number of loops, $I$ the total number of internal lines, $V_1$ the number of ghost-gluon vertices, $V_3$ the number of three-gluon vertices, $V_4$ the number of four-gluon vertices, $E$ the total number of external legs, $E_c$ the number of ghost or antighost legs. 

The superficial degree of divergence writes
\beq
\delta=dL-2I+V_1+V_3-E_c\,,
\eeq 
where $\smash{V_1+V_3}$ corresponds to the additional momenta coming from the derivative vertices and $-E_c$ accounts for the fact that each external ghost or antighost leg brings a factor that carries only the external momentum (and thus does not count in the evaluation of $\delta$). Momentum conservation implies that
\beq
L=I-(V_1+V_3+V_4)+1\,.
\eeq
Moreover, the number of internal and external lines is related to the number of vertices of each type as
\beq
E+2I=3(V_1+V_3)+4V_4\,.
\eeq
We then have 
\beq
\delta & \!\!\!=\!\!\! &  d+(d-2)I-(d-1)(V_1+V_3)-dV_4-E_c\nonumber\\
& \!\!\!=\!\!\! & d+\frac{d-2}{2}(3(V_1+V_3)+4V_4-E)-(d-1)(V_1+V_3)-dV_4-E_c\nonumber\\
& \!\!\!=\!\!\! & d+\frac{d-4}{2}(V_1+V_3+2V_4)-\frac{d-2}{2}E-E_c\,.
\eeq
In $d=4$ dimensions, this becomes
\beq
\boxed{\delta=4-E-E_c}\,\,.\label{eq:delta}
\eeq
This shows that the divergences are limited to a finite number of primitively divergent functions.

Note that the above formula for $\delta$ can be understood and even derived from simple dimensional analysis considerations. Indeed, unless there are some specific symmetries, the power of internal momenta of a given diagram is equal to the dimension of that diagram minus the mass dimension of the couplings multiplying that diagram. The dimension of a correlation function with $E$ external legs in configuration space is $E(d-2)/2$ but, in Fourier space, taking into account momentum conservation, this gives the dimension $E(d-2)/2-Ed+d$. And finally, when constructing the 1PI vertex, one needs to amputate $E$ external propagators of dimension $-2$ each. So the dimension of the 1PI vertex with $E$ external legs is
\beq
\frac{d-2}{2}E-dE+d+2E=d-\frac{d-2}{2}E\,.
\eeq
From this, we need to remove the mass dimension of the couplings in front of the diagram. Since the dimension of the coupling is $(4-d)/2$, we arrive at
\beq
d-\frac{d-2}{2}E+\frac{d-4}{2}(V_1+V_3+2V4)\,,
\eeq
which is essentially the formula above, up to the fact, that because we may have extra factors of external momenta associated to external ghost or antighost legs, we need also to subtract those in the determination of the power of internal momenta. This explains the additional $-E_c$.

\section{Symmetry Constraints}
In order to show that the CF action is renormalizable, we need to show that the divergences that appear in the primitively divergent functions can be absorbed in operators that can be interpreted as mere redefinitions of the fields and parameters of the original model. 

There are many hints that point in that direction, as can be inferred from the very term $-E_c$ in Eq.~(\ref{eq:delta}). In general, the superficial degree of divergence of a given vertex involving ghost and antighost legs, is suppressed by as many power of momenta as there are ghost and antighost legs, compared to the corresponding vertex where are all the ghost and antighost legs are replaced by gluons legs. For instance, the superficial degrees of divergence of two-ghost/two-antighost vertex and of the four-gluon vertex differ by $-4$. Because the four-gluon vertex is logarithmically divergent, this means that the two-ghost/two-antighost vertex is finite and does not require the introduction of a quartic ghost interaction terms in the action that would modify the original CF model.

But this is not enough to fix the form of the counterterm action, specially in the gauge-field sector. To do so, one needs to exploit the symmetries of the model. If the symmetries are strong enough, they might constraint the vertex functions and thus their divergences and  corrections to be added to the action to have precisely the same form as the original action. The way to proceed in general to write these constraints is to perform a change of variables under the functional integral that defines the partition function, in the form of the considered symmetry:
\beq
Z[J] & \!\!\!=\!\!\! & \int {\cal D}\varphi\,e^{-S[\varphi]+\int_x J(x)\varphi(x)}\nonumber\\
& \!\!\!=\!\!\! & \int {\cal D}[\varphi+\delta\varphi]\,e^{-S[\varphi+\delta\varphi]+\int_x J(x)(\varphi(x)+\delta\varphi(x))}\,.
\eeq
In the case where both the action and the functional integral measure are invariant, we deduce that
\beq
0=\int {\cal D}\varphi\,e^{-S[\varphi]+\int_x J(x)\varphi(x)}\int_x J(x)\delta\varphi(x)\,,
\eeq
and, by taking functional derivatives with respect to the source, one obtains constraints on the vertex functions. This is quite nice in the case of linear symmetries. However, for non-linear transformations such as those in the mBRST symmetry, the so-obtained equations contain insertions of composite operators which is not our primary goal here. 

To cope with this, one can exploit the nilpotency of $s$ in the Faddeev-Popov case or the simple properties of $\hat s^2$ in the Curci-Ferrari case. More precisely, one includes additional sources for the transformations of $A_\mu^a$ and $c^a$:
\beq
Z[J,\eta,\bar\eta,R;\bar K,\bar L]\equiv \int{\cal D}[Ac\bar ch]\,e^{-S+\int_x(J_\mu^aA_\mu^a+\bar\eta^ac^a+\bar c^a\eta^a+R^aih^a+\bar K_\mu^a\hat sA_\mu^a+\bar L^a\hat sc^a)}\,.\nonumber\\\label{eq:ZK}
\eeq
Then upon performing a change of variables in the form of a mBRS transformation and using that $\hat s$ remains nilpotent in the $A-c$ sector, one finds
\beq
0 & = & \int{\cal D}[Ac\bar ch]\,e^{-S+\int_x(J_\mu^aA_\mu^a+\bar\eta^ac^a+\bar c^a\eta^a+R^aih^a+\bar K_\mu^a\hat sA_\mu^a+\bar L^a\hat sc^a)}\nonumber\\
& & \times \int_x(J_\mu^a \hat{s}A_\mu^a-\bar\eta^a \hat{s}c^a+\hat{s}\bar c^a\eta^a+R^a\hat{s}ih^a)\,.
\eeq
The beauty of it is that the non-linear terms can now be expressed as derivative of $Z$ or $W$ with respect to the newly introduced sources. In fact, one finds
\beq
0=\int_x \left\{J_\mu^a\frac{\delta W}{\delta\bar K_\mu^a}-\bar\eta^a\frac{\delta W}{\delta\bar L^a}+\frac{\delta W}{\delta R^a}\eta^a+m^2R^a\frac{\delta W}{\delta\bar\eta^a}\right\}.
\eeq
It is more convenient to rewrite this equation in terms of the effective action $\Gamma$. Since the sources $\bar K_\mu^a$ and $\bar L^a$ do not enter the Legendre transform, they remain variables of $\Gamma$, with
\beq
\frac{\delta W}{\delta\bar K_\mu^a(x)}=-\frac{\delta \Gamma}{\delta\bar K_\mu^a(x)} \quad {\rm and} \quad \frac{\delta W}{\delta\bar L^a(x)}=-\frac{\delta \Gamma}{\delta\bar L^a(x)}\,,
\eeq
in addition to (\ref{eq:sources1})-(\ref{eq:sources2}). The symmetry constraint reads eventually
\beq
\boxed{0=\int_x \left\{\frac{\delta\Gamma}{\delta A_\mu^a}\frac{\delta\Gamma}{\delta\bar K_\mu^a}+\frac{\delta\Gamma}{\delta c^a}\frac{\delta \Gamma}{\delta\bar L^a}-ih^a\frac{\delta\Gamma}{\delta \bar c^a}-m^2c^a\frac{\delta\Gamma}{\delta ih^a}\right\}}\,\,.
\eeq
This is known as the Zinn-Justin equation.

To get some intuition on the information contained in this equation, let us notice that it should be verified by the tree-level contribution to $\Gamma$ given by
\beq
\Gamma_{\rm tree}=S-\int_x (\bar K_\mu^a\hat sA_\mu^a+\bar L^a\hat sc^a)\,.
\eeq
Plugging this form in the Zinn-Justin equation, we find
\beq
0=\int_x \left\{\frac{\delta\Gamma_{\rm tree}}{\delta A_\mu^a}\hat sA_\mu^a+\frac{\delta\Gamma_{\rm tree}}{\delta c^a}\hat sc^a+ih^a\frac{\delta\Gamma_{\rm tree}}{\delta \bar c^a}+m^2c^a\frac{\delta\Gamma_{\rm tree}}{\delta ih^a}\right\}.
\eeq
Defining $\hat s\bar c^a$ and $\hat sih^a$ as above, this becomes
\beq
0=\int_x \left\{\hat sA_\mu^a\frac{\delta\Gamma_{\rm tree}}{\delta A_\mu^a}+\hat sc^a\frac{\delta\Gamma_{\rm tree}}{\delta c^a}+\hat s\bar c^a\frac{\delta\Gamma_{\rm tree}}{\delta \bar c^a}+\hat sih^a\frac{\delta\Gamma_{\rm tree}}{\delta ih^a}\right\}.
\eeq
Collecting together those terms that depend on a specific source, and those that do not, we obtain three constraints
\beq
0 & \!\!\!=\!\!\! & \int_x \left\{\hat sA_\mu^a\frac{\delta S}{\delta A_\mu^a}+\hat sc^a\frac{\delta S}{\delta c^a}+\hat s\bar c^a\frac{\delta S}{\delta \bar c^a}+\hat sih^a\frac{\delta S}{\delta ih^a}\right\},\\
0 & \!\!\!=\!\!\! & \int_x \left\{\hat sA_\mu^a\frac{\delta\hat sA_\nu^b}{\delta A_\mu^a}+\hat sc^a\frac{\delta\hat sA_\nu^b}{\delta c^a}+\hat s\bar c^a\frac{\delta\hat sA_\nu^b}{\delta \bar c^a}+\hat sih^a\frac{\delta\hat sA_\nu^b}{\delta ih^a}\right\},\\
0 & \!\!\!=\!\!\! & \int_x \left\{\hat sA_\mu^a\frac{\delta\hat sc^b}{\delta A_\mu^a}+\hat sc^a\frac{\delta\hat sc^b}{\delta c^a}+\hat s\bar c^a\frac{\delta\hat sc^b}{\delta \bar c^a}+\hat sih^a\frac{\delta\hat sc^b}{\delta ih^a}\right\}.
\eeq
The first of them encodes the invariance of the action under mBRS, while the other two encode the nilpotency of the mBRS transformation in the $A-c$ sector.

\section{The Structure of UV Divergences}
Let us now see how the mBRST symmetry constrains the divergences. The Zinn-Justin equation should hold for the divergent part of $\Gamma$. But the latter being local with operators up to and including dimension $4$, it should write
\beq
\Gamma_{\rm div}=\Gamma_{\rm div}[A,c,\bar c,ih]+\int_x (\bar K_\mu^a\tilde sA_\mu^a+\bar L^a\tilde sc^a)\,,
\eeq
where $\tilde sA_\mu^a$ and $\tilde sc^a$ are dimension $2$ functionals with ghost number $+1$ and $+2$ respectively. Plugging this in the Zinn-Justin equation, while setting $\tilde s\bar c^a=ih^a$ and $\tilde sih^a=m^2c^a$, we deduce that $\tilde s$ is nilpotent in the sector $A-c$ and leaves $\Gamma_{\rm div}[A,c,\bar c,ih]$ invariant.

From this, it is not difficult to show that
\beq
\tilde sc^a & \!\!\!=\!\!\! & -\frac{X}{2}gf^{abc} c^b c^c\,,\\
\tilde sA_\mu^a & \!\!\!=\!\!\! & Y\partial_\mu c^a+Xgf^{abc}A_\mu^b c^c\,.
\eeq
Let us then express the invariance of $\Gamma_{\rm div}[A,c,\bar c,ih]$ under $\tilde s$. We can use the fact that the Nakanishi-Lautrup sector is not renormalized and the fact that the antighost comes always with a derivative to write\footnote{In fact, from the analysis in the main text, we know already that the correction term to the ghost-antighost-gluon coupling is not needed from the non-renormalizaiton theorem. It is interesting however to do the analysis without any knowledge of the non-renormalization theorems and to rederive them from this more general analysis. We stress however, that the two non-renormalization theorems can be derived in a simpler way without using the Zinn-Justin equation, as we have explained in the main text.}
\beq
\Gamma_{\rm div}[A,c,\bar c,ih]=\Gamma_{\rm div}[A]+\int_x\left\{\frac{1}{Z_c}\partial_\mu\bar c^a\partial_\mu c^a+gC^{abc}\partial_\mu\bar c^a A_\mu^b c^c+ih^a\partial_\mu A_\mu^a\right\}.\nonumber\\
\eeq
The variation under $\tilde s$ is
\beq
0 & \!\!\!=\!\!\! & \tilde s\Gamma_{\rm div}[A,c,\bar c,ih]\nonumber\\
& \!\!\!=\!\!\! & \tilde s\Gamma_{\rm div}[A]\nonumber\\
& \!\!\!+\!\!\! & \int_x\Bigg\{\frac{1}{Z_c}\partial_\mu ih^a\partial_\mu c^a-\frac{X}{Z_c}gf^{abc}\partial_\mu\bar c^a c^b\partial_\mu c^c+gC^{abc}\partial_\mu ih^a A_\mu^bc^c\nonumber\\
& \!\!\!-\!\!\! & gC^{abc}\partial_\mu\bar c^a(Y\partial_\mu c^b+Xgf^{bde}A_\mu^dc^e)c^c+\frac{X}{2}g^2C^{abc}f^{cde}\partial_\mu\bar c^aA_\mu^bc^dc^e\nonumber\\
& \!\!\!-\!\!\! & \partial_\mu ih^a(Y\partial_\mu c^a+Xgf^{abc}A_\mu^b c^c)+m^2 c^a\partial_\mu A_\mu^a\Bigg\}.
\eeq
Since $\tilde s\Gamma_{\rm div}[A]$ contains neither $h$ nor $\bar c$, let us first look at those terms involving these fields. We get
\beq
Y=\frac{1}{Z_c}\,, \quad C^{abc}=Xf^{abc}\,,
\eeq
and we are left with
\beq
0=\tilde s\Gamma_{\rm div}[A,c,\bar c,ih] & \!\!\!=\!\!\! & \tilde s\Gamma_{\rm div}[A]+\int_x m^2 c^a\partial_\mu A_\mu^a\nonumber\\
& \!\!\!=\!\!\! & \tilde s\left(\Gamma_{\rm div}[A]-\int_x\frac{m^2}{2Y}A_\mu^aA_\mu^a\right).
\eeq
We deduce that
\beq
\hat\Gamma_{\rm div}[A,c,\bar c,ih]=\int_x\left(\frac{1}{4Z_A}F^{'a}_{\mu\nu}F^{'a}_{\mu\nu}+\frac{1}{Z_c}\partial_\mu\bar c^a{\cal D}'c^a+ih^a\partial_\mu A_\mu^a+\frac{m^2}{2Y}A_\mu^aA_\mu^a\right),\nonumber\\
\eeq
with
\beq
{\cal D}'_\mu\varphi^a & \!\!\!=\!\!\! & \partial_\mu\varphi^a+\frac{X}{Y}gf^{abc}A_\mu^b \varphi^c\,,\\
F^{'a}_{\mu\nu} & \!\!\!=\!\!\! & \partial_\mu A_\nu^a-\partial_\nu A_\mu^a+\frac{X}{Y}gf^{abc}A_\mu^bA_\nu^c\,.
\eeq
The divergences can be absorbed in a redefinition of the fields $A\to\sqrt{Z_A}A$, $c\to Z_c\,c$ and the parameters $g\to Z_g g$, $m\to Z_m m$ with $Z_g\sqrt{Z_A}X/Y$ and $Z_m^2Z_A/Y$ both finite. Note that because, $\tilde sc^a\propto X$, its divergence can be absorbed in the source $\bar L^a$ and thus, without loss of generality, we can assume that $\smash{X=1}$. Moreover, since $Y=1/Z_c$, we find that both $Z_g\sqrt{Z_A}Z_c$ and $Z_{m^2}Z_AZ_c$ are finite. Note also that the effective action in the presence of the sources $\bar K_\mu^a$ and $\bar L^a$ is also finite provided the latter sources get the renormalized by the respective factors $Z_c^{1/2}$ and $Z_A^{1/2}$ respectively. It can then be checked that the Zinn-Justin equation in terms of renormalized fields and sources keeps the same form with the bare square mass replaced by $Z_A Z_c Z_{m^2} m^2$ where $m^2$ is the renormalized mass and $Z_A Z_c Z_{m^2}$ is a finite combination that can be taken equal to $1$ as part of the definition of the renormalization scheme.

For another way of deriving the non-renormalization theorem for the mass renormalization factor (other than the one presented in this Appendix or the one presented in the main test) and that combines the Zinn-Justin equation with the quantum equations of motion, see \cite{Tissier:2010ts}.

\chapter{Master Integrals}

Let us here gather the master integrals needed for the evaluation of the one-loop ghost and gluon self-energies.

\section{The Tadpole Integral}
The tadpole integrals writes
\beq
J_m & \!\!\!=\!\!\! & \int_Q \frac{1}{Q^2+m^2}=\frac{(m^2)^{d/2-1}}{(4\pi)^{d/2}}\Gamma(1-d/2)\nonumber\\
& \!\!\!=\!\!\! & -\frac{m^2}{16\pi^2}\left[\frac{1}{\epsilon}+\ln\frac{\bar\mu^2}{m^2}+1+{\cal O}(\epsilon)\right]\,,
\eeq
where $\smash{\bar\mu^2\equiv 4\pi\mu^2e^{-\gamma}}$ and $\gamma$ is the Euler-Mascheroni constant. The $\mu$ appears because each integral is originally multiplied by a bare $g_B^2$ which in dimensional regularization is $\mu^{2\epsilon}g^2Z_{g^2}$ in terms of the renormalized coupling. The factor $\mu^{2\epsilon}$ is then absorbed in a redefinition of $\int_Q$.

\section{The Bubble Integral}
The bubble integral writes
\beq
I_{m_1m_2}(P) & \!\!\!=\!\!\! & \int_Q \frac{1}{(Q^2+m^2_1)((Q+P)^2+m^2_2)}\nonumber\\
& \!\!\!=\!\!\! & \int_0^1dx \int_Q \frac{1}{(Q^2+xm^2_1+(1-x)m^2_2+x(1-x)P^2)^2}\nonumber\\
& \!\!\!=\!\!\! & \frac{\Gamma(2-d/2)}{(4\pi)^{d/2}}\int_0^1dx (xm^2_1+(1-x)m^2_2+x(1-x)P^2)^{d/2-2}\nonumber\\
& \!\!\!=\!\!\! & \frac{\Gamma(\epsilon)}{(4\pi)^{2-\epsilon}}\int_0^1dx (xm^2_1+(1-x)m^2_2+x(1-x)P^2)^{-\epsilon}\nonumber\\
& \!\!\!=\!\!\! & \frac{1}{16\pi^2}\left[\frac{1}{\epsilon}-\int_0^1dx \ln \frac{xm^2_1+(1-x)m^2_2+x(1-x)P^2}{\bar\mu^2}+{\cal O}(\epsilon)\right]\,.\nonumber\\
\eeq
Although the integral can be computed in full generality, let us here only consider the cases that are needed in the CF model
\beq
I_{m0}(P) & \!\!\!=\!\!\! & \frac{1}{16\pi^2}\left[\frac{1}{\epsilon}-\int_0^1 dx\,\ln x-\int_0^1dx \ln \frac{m^2+(1-x)P^2}{\bar\mu^2}+{\cal O}(\epsilon)\right]\nonumber\\
& \!\!\!=\!\!\! & \frac{1}{16\pi^2}\left[\frac{1}{\epsilon}+\ln\frac{\bar\mu^2}{P^2}+2+\frac{m^2}{P^2}\ln\frac{m^2}{P^2}-\left(1+\frac{m^2}{P^2}\right)\ln\left(1+\frac{m^2}{P^2}\right)+{\cal O}(\epsilon)\right].\nonumber\\
\eeq
We can test this formula by taking the $\smash{P\to 0}$ limit since
\beq
I_{m0}(0)=\frac{J_0-J_m}{m^2}=-\frac{J_m}{m^2}=\frac{1}{16\pi^2}\left[\frac{1}{\epsilon}+\ln\frac{\bar\mu^2}{m^2}+1+{\cal O}(\epsilon)\right]\,.
\eeq
This is what we find indeed by taking the limit in the explicit expression above. Notice also that by taking the $\smash{m\to 0}$ limit, we find
\beq
I_{00}(P) & = & \frac{1}{16\pi^2}\left[\frac{1}{\epsilon}+\ln\frac{\bar\mu^2}{P^2}+2+{\cal O}(\epsilon)\right].
\eeq
Finally, we consider
\beq
I_{mm}(P) & \!\!\!=\!\!\! & \frac{1}{16\pi^2}\left[\frac{1}{\epsilon}-\int_0^1dx \ln \frac{m^2+x(1-x)P^2}{\bar\mu^2}+{\cal O}(\epsilon)\right]\nonumber\\
& \!\!\!=\!\!\! & \frac{1}{16\pi^2}\left[\frac{1}{\epsilon}-\int_{-1/2}^{1/2}dy \ln \frac{m^2+P^2/4-P^2y^2}{\bar\mu^2}+{\cal O}(\epsilon)\right]\nonumber\\
& \!\!\!=\!\!\! & \frac{1}{16\pi^2}\left[\frac{1}{\epsilon}+\ln\frac{\bar\mu^2}{P^2}-\int_{-1/2}^{1/2}dy \ln \left(m^2/P^2+1/4-y^2\right)+{\cal O}(\epsilon)\right]\nonumber\\
& \!\!\!=\!\!\! & \frac{1}{16\pi^2}\left[\frac{1}{\epsilon}+\ln\frac{\bar\mu^2}{P^2}-2\int_{-1/2}^{1/2}dy \ln \left(\sqrt{\frac{m^2}{P^2}+\frac{1}{4}}+y\right)+{\cal O}(\epsilon)\right]\nonumber\\
& \!\!\!=\!\!\! & \frac{1}{16\pi^2}\left[\frac{1}{\epsilon}+\ln\frac{\bar\mu^2}{P^2}\right.\nonumber\\
& & -\,2\left(\sqrt{\frac{m^2}{P^2}+\frac{1}{4}}+1/2\right)\left[\ln \left(\sqrt{\frac{m^2}{P^2}+\frac{1}{4}}+1/2\right)-1\right]\nonumber\\
& &  \left.+\,2 \left(\sqrt{\frac{m^2}{P^2}+\frac{1}{4}}-1/2\right)\left[\ln \left(\sqrt{\frac{m^2}{P^2}+\frac{1}{4}}-1/2\right)-1\right]+{\cal O}(\epsilon)\right]\nonumber\\
& \!\!\!=\!\!\! & \frac{1}{16\pi^2}\left[\frac{1}{\epsilon}+\ln\frac{\bar\mu^2}{m^2}+2+\sqrt{1+\frac{4m^2}{P^2}}\ln \frac{\sqrt{1+\frac{4m^2}{P^2}}-1}{\sqrt{1+\frac{4m^2}{P^2}}+1}+{\cal O}(\epsilon)\right].\nonumber\\
\eeq

\section{Final expressions for the self-energies}
Using the above expressions for $J_m$, $I_{mm}$, $I_{m0}$ and $I_{00}$ in Eqs.~(\ref{eq:248}) and (\ref{eq:264}), the final expressions for the one-loop self-energies within the CF model are
\beq
\frac{\Sigma(P)}{g^2N} & \!\!\!=\!\!\! & -\frac{3P^2}{64\pi^2}\left(\frac{1}{\epsilon}+\ln\frac{\bar\mu^2}{m^2}\right)-\frac{5P^2+m^2}{64\pi^2}\nonumber\\
& \!\!\!+\!\!\! & \frac{1}{64\pi^2}\left[\frac{(P^2+m^2)^3}{P^2m^2}\ln\left(1+\frac{P^2}{m^2}\right)-\frac{(P^2)^2}{m^2}\ln\frac{P^2}{m^2}\right],
\eeq
and
\beq
\frac{\Pi^\perp(P)}{g^2N} & \!\!\!=\!\!\! & \frac{1}{64\pi^2}\left(\frac{1}{\epsilon}+\frac{2}{3}+\ln\frac{\bar\mu^2}{m^2}\right)\left(-\frac{26}{3}P^2+3m^2\right)+\frac{-23P^2+57m^2}{192\pi^2}\nonumber\\
& \!\!\!+\!\!\! & \frac{P^2}{384\pi^2}\left[\left(1+\frac{4m^2}{P^2}\right)^{3/2}\left(\left(\frac{P^2}{m^2}\right)^2-20\frac{P^2}{m^2}+12\right)\ln\frac{\sqrt{1+\frac{4m^2}{P^2}}-1}{\sqrt{1+\frac{4m^2}{P^2}}+1}\right.\nonumber\\
& & \hspace{1.5cm}+\,2\left(1+\frac{m^2}{P^2}\right)^3\left(\left(\frac{P^2}{m^2}\right)^2-10\frac{P^2}{m^2}+1\right)\ln\left(1+\frac{P^2}{m^2}\right)\nonumber\\
& & \hspace{1.5cm}+\,\left.\left(2-\left(\frac{P^2}{m^2}\right)^2\right)\ln\frac{P^2}{m^2}-2\left(\frac{m^2}{P^2}\right)^2\right].
\eeq

\end{document}